\documentclass[letterpaper,twocolumn,11pt,accepted=2025-08-06]{quantumarticle}

\pdfoutput=1
\usepackage[main=english,german,ngerman]{babel}
\usepackage{xcolor,colortbl}
\usepackage{braket}
\usepackage{dsfont}
\usepackage{hyphenat}
\usepackage[numbers,sort&compress]{natbib}
\usepackage{array}
\usepackage[export]{adjustbox}
\usepackage{amsmath,amssymb,amsfonts}
\usepackage[export]{adjustbox}
\usepackage{graphicx,wrapfig,lipsum}
\usepackage{dcolumn}
\usepackage{bm}
\usepackage{siunitx}
\usepackage{enumitem} 
\usepackage{booktabs}
\usepackage{mathtools}
\usepackage[most]{tcolorbox}
\usepackage{caption}
\usepackage{subcaption}
\usepackage[colorlinks=true,linkcolor=magenta,urlcolor=magenta,citecolor=magenta,breaklinks=true]{hyperref}
\hypersetup{colorlinks}
\usepackage[capitalise]{cleveref}
\usepackage{physics}
\usepackage{bbold}
\usepackage{float}
\usepackage{multirow}
\usepackage[utf8]{inputenc}
\usepackage[toc,page]{appendix}
\usepackage{xspace}
\usepackage{suffix}
\usepackage{xstring}
\usepackage[nolist, nohyperlinks]{acronym}
\usepackage{comment}
\usepackage{makecell}

\newcommand{\mc}[1]{\ensuremath{\mathcal{#1}}}

\newcommand{\Riegrad}{\ensuremath{\grad_{\text{Rie}}}}
\newcommand{\ind}[1]{\ensuremath{\text{ind}(#1)}}

\newcommand{\Uref}{\ensuremath{U_\text{ref}}}

\newcommand{\topenv}{\ensuremath{E_\text{top}}}
\newcommand{\bottomenv}{\ensuremath{E_\text{bottom}}}
\newcommand{\leftenv}{\ensuremath{E_\text{left}}}
\newcommand{\rightenv}{\ensuremath{E_\text{right}}}

\newcommand{\overlap}{\ensuremath{\mathcal{T}(W)}}
\newcommand{\loss}{\ensuremath{\mathcal{L}}}
\newcommand{\cost}{\ensuremath{\mathcal{C}}}
\newcommand{\costF}{\ensuremath{\mathcal{C}_{F}}}
\newcommand{\costHS}{\ensuremath{\mathcal{C}_{HS}}}
\newcommand{\maxbondim}{\ensuremath{\chi_\text{max}}}
\newcommand{\Trotter}[2]{\ensuremath{U_\text{Trotter}^{\text{#1} #2}}}

\newcommand{\C}{\ensuremath{\mathds{C}}}
\newcommand{\R}{\ensuremath{\mathds{R}}}

\newcommand{\errorThres}{\ensuremath{\epsilon_\text{thres}}}
\newcommand{\errorFinal}{\ensuremath{\mc{C}_\text{final}}}

\newcommand{\errorTrotter}{\ensuremath{\varepsilon_\text{Trot}}}
\newcommand{\errorTruncation}{\ensuremath{\varepsilon_\text{trunc}}}
\newcommand{\errorCompression}{\ensuremath{\varepsilon_\text{comp}}}

\begin{document}

\title{Riemannian quantum circuit optimization based on matrix product operators}

\author{Isabel~Nha~Minh~Le}
\orcid{0000-0001-6707-044X}
\email{isabel.le@tum.de}
\affiliation{Technical University of Munich, School of Computation, Information and Technology, Boltzmannstr. 3, 85748 Garching, Germany}
\affiliation{Munich Center for Quantum Science and Technology (MCQST), Schellingstr. 4, 80799 Munich, Germany}
\author{Shuo~Sun}
\orcid{0009-0006-5775-9730}
\affiliation{Technical University of Munich, School of Computation, Information and Technology, Boltzmannstr. 3, 85748 Garching, Germany}
\affiliation{Munich Center for Quantum Science and Technology (MCQST), Schellingstr. 4, 80799 Munich, Germany}
\author{Christian~B.~Mendl} 
\orcid{0000-0002-6386-0230}
\email{christian.mendl@tum.de}
\affiliation{Technical University of Munich, School of Computation, Information and Technology, Boltzmannstr. 3, 85748 Garching, Germany}
\affiliation{Munich Center for Quantum Science and Technology (MCQST), Schellingstr. 4, 80799 Munich, Germany}
\affiliation{Technical University of Munich, Institute for Advanced Study, Lichtenbergstr. 2a, 85748 Garching, Germany}

\begin{abstract}
We significantly enhance the simulation accuracy of initial Trotter circuits for Hamiltonian simulation of quantum systems by integrating first-order Riemannian optimization with tensor network methods. 
Unlike previous approaches, our method imposes no symmetry assumptions, such as translational invariance, on the quantum systems. This technique is scalable to large systems through the use of a matrix product operator representation of the reference time evolution propagator. 
Our optimization routine is applied to various spin chains and fermionic systems described by the transverse-field Ising Hamiltonian, the Heisenberg Hamiltonian, and the spinful Fermi-Hubbard Hamiltonian. 
In these cases, our approach achieves a relative error improvement of up to four orders of magnitude for systems of 50 qubits, although our method is also applicable to larger systems. 
Furthermore, we demonstrate the versatility of our method by applying it to molecular systems, specifically lithium hydride, achieving an error improvement of up to eight orders of magnitude. 
This proof of concept highlights the potential of our approach for broader applications in quantum simulations.
\end{abstract}

\maketitle

\section{Introduction}

Simulating the time evolution of a quantum system was the original motivation behind quantum computers and remains one of their most natural applications~\cite{feynman2018simulating,lloyd1996universal,zalka1998simulating}, particularly in quantum chemistry and quantum many-body physics~\cite{bauer2020quantum,cao2019quantum,eisert2015quantum}. 
A widely used method for this purpose is Trotterization~\cite{kluber2023trotterization,childs2021theory}, where the time propagator $U_t = e^{-iHt}$ is approximated using lower-body operators. 
This approach is straightforward to implement and comes in various orders, each with different circuit depths and error scalings~\cite{childs2021theory,schubert2023trotter}. 
However, achieving high accuracy for long-time simulations necessitates deep quantum circuits. 
This raises some interesting questions: 
\begin{itemize}[label=$\circ$] 
    \item Can the same quantum circuit layout provided by Trotterization achieve better accuracy? 
    \item And is it possible to optimize the quantum gates using only classical computation so that expensive quantum resources are only needed for the (optimized) Hamiltonian simulation? 
\end{itemize} 
These questions have been previously explored by (i)~considering specific parameterizations of the quantum gates~\cite{mansuroglu2023variational,mansuroglu2023problem,tepaske2023optimal,mc2023classically,mc2024towards,robertson2023approximate,zhang2024scalable}, and (ii)~directly interpreting the quantum gates as unitary matrices in a brickwall circuit~\cite{kotil2024riemannian,causer2024scalable,gibbs2024deep,anselme2024combining}. 
In this work, we focus on the latter approach and thereby circumvent possible dependencies of the optimization on the chosen parameterization. 

In Ref.~\cite{kotil2024riemannian}, Riemannian optimization and tensor network methods are combined to simultaneously optimize all quantum gates under the constraint of unitarity for small translationally-invariant quantum systems using an explicit full-rank matrix reference of $U_t$. 
Due to the translational invariance, the same optimized quantum gates can simulate the quantum dynamics of corresponding larger systems. 
In contrast, Refs.~\cite{causer2024scalable,gibbs2024deep} employ a local quantum gate update inspired by tensor network methods to directly optimize larger quantum systems. 
To this end, they approximate $U_t$ using a fine Trotterization for a suitable time step $t$, allowing it to be expressed efficiently as a \ac{MPO}. 
The classically optimized quantum circuit can then repeatedly be executed on quantum hardware to enable the simulation of quantum dynamics for longer times $t'\gg t$. 

This work introduces a novel alternative quantum circuit optimization approach by integrating the ideas and methods of Refs.~\cite{kotil2024riemannian,causer2024scalable,gibbs2024deep}. 
Specifically, we focus on quantum systems where a Trotter step can be represented as a brickwall circuit and introduce a simultaneous-update optimization method to enhance quantum gates under the unitarity constraint. 
To achieve this, we adapt the gradient-based ADAM optimizer~\cite{diederik2014adam} to the framework of Riemannian optimization on the complex Stiefel manifold, i.e., the manifold of unitary matrices. 
We approximate the time evolution reference operator $U_t$ to target larger quantum systems by employing a higher-order Trotterization \ac{MPO} and evaluate the cost function and corresponding gradient utilizing tensor network methods. 
Our Riemannian optimization approach updates the quantum gates of the quantum circuit simultaneously in contrast to the local gate updates used in Refs.~\cite{causer2024scalable,gibbs2024deep}. This also leads to a potential promotion to second-order Riemannian optimization -- something that is not possible in the context of local gate updates.
Furthermore, the presented approach does not assume any symmetry invariance of the quantum system to be simulated as in Ref.~\cite{kotil2024riemannian}. 

While numerous scalable methods for quantum circuit compression have been proposed~\cite{mansuroglu2023variational,mansuroglu2023problem,tepaske2023optimal,mc2023classically,mc2024towards,robertson2023approximate,zhang2024scalable,kotil2024riemannian,causer2024scalable,gibbs2024deep,anselme2024combining}, our approach distinguishes itself by framing the problem within the context of differential geometry. This geometric perspective not only facilitates efficient optimization but also potentially enables access to intrinsic, hardware-independent measures of circuit complexity~\cite{dowling2006geometry,nielsen2006quantum}. Notably, the shortest path between unitaries under a Riemannian metric can be defined via the geodesic equation, offering a principled benchmark for circuit optimality. Such geometric tools can also be used in potential alternative optimization strategies~\cite{lewis2025geodesic}. Importantly, two circuits may yield comparable fidelity yet differ substantially in their geometric complexity -- a distinction our Riemannian framework could, in principle, capture, unlike methods that do not account for the manifold structure of the unitary group.

We apply the presented method to exemplary spin chains and fermionic systems. 
Specifically, we conduct numerical experiments on systems of 50 qubits for a non-integrable case of the Ising Hamiltonian and a Heisenberg model, explicitly considering non-translationally invariant systems. 

For fermionic systems, we utilize the fermionic swap network~\cite{kivlichan2018quantum} and optimize a one-dimensional spinful Fermi-Hubbard model of 50 spin orbitals, as well as a paradigmatic example of \ac{LiH} for the molecular Hamiltonian. 

\paragraph*{Related research.}
Concurrently with our research, two independent studies have been published that combine the Riemannian ADAM optimizer, automatic differentiation, and tensor network methods to optimize a brickwall circuit~\cite{rogerson2024quantum,guo2025efficient}. 
The first study optimizes a quantum state represented as a matrix product state, while we directly optimize the time evolution operator, allowing for an arbitrary choice of initial quantum state. 
The second work explicitly compiles the brickwall circuit into single-qubit and CNOT gates. 
In contrast, we view the compilation as an excluded step and do not specify a fixed native gate set. 
Furthermore, we analytically implement the required gradient without relying on additional software packages for automatic differentiation. \\

The remaining paper is structured as follows: 
We first introduce the optimization problem in \cref{sec:setup}, by presenting the quantum circuit layout, defining the cost function and its gradient, formulating the underlying optimization problem, and introducing the utilized initialization methods. 
We then give a brief recap on Riemannian optimization and adapt the ADAM optimizer to the relevant case of the complex Stiefel manifold in \cref{sec:riemannian}. 
Furthermore, we present the tensor network methods utilized to efficiently represent the unitary reference and to evaluate the cost function and its gradient in \cref{sec:tn-methods}. We additionally point out how these methods can be used within an alternative parameterization-based approach.
In \cref{sec:results}, we show the results of the conducted numerical simulations for various (i)~spin chains, and (ii)~fermionic systems. 
Finally, we give a conclusion and outlook. 
\section{Optimization problem \label{sec:setup}}
\paragraph*{Quantum~circuit~layout.}
We consider a so-called brickwall circuit layout built from $L$ layers of two-qubit gates, where the $\ell$-th layer of the quantum circuit $W^\ell$ has $n_\ell$ gates. 
We denote the $i$-th gate in layer $W^\ell$ as $G_i^\ell$ and the concatenation of layers $i$ to $j$ as $W^{i:j}$, i.e.,
\[
W^\ell = \bigotimes_{i=1}^{n_\ell} G_i^\ell \,\,\, \text{and} \,\,\, W^{i:j} = \prod_{\ell=i}^j W^\ell.
\]
Finally, we call the overall brickwall circuit $W=W^{1:L}$. 

\paragraph*{Cost function.}
A common way~\cite{kotil2024riemannian, causer2024scalable} to quantify how well a quantum circuit $W$ of $N$ qubits approximates a reference unitary $\Uref$ is given by the normalized Frobenius norm:
\begin{align*}
    \costF(\Uref,W) &= \frac{1}{2d}||\Uref-W||_F^2 \\
    &= \frac{1}{2d}\Tr\left( [\Uref-W]^\dagger [\Uref-W] \right) \\
    &= 1 - \frac{1}{d}\Re\Tr\left(\Uref^\dagger W\right),
\end{align*}
where $d=2^N$. 
The trace $\Tr(\Uref^\dagger W)$ is the overlap between the reference unitary and a brickwall circuit and plays a crucial role in the presented method. 
Using the tensor networks notation, this overlap can be diagrammatically expressed as 
\begin{align}\label{eq:overlap}
    \overlap &=\Tr(\Uref^\dagger W) \nonumber \\
    &= 
  \begin{minipage}[h]{5.2cm}
	\vspace{0pt}
	\includegraphics[width=5.2cm]{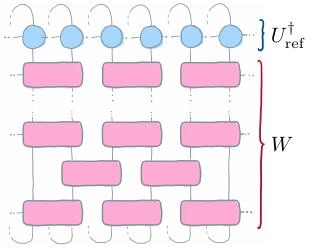}
   \end{minipage}
\end{align}
 
%However, the Frobenius norm is sensitive to global phases, such that it is more restrictive than required in practice. 
The Frobenius norm is sensitive to global phases, rendering it more restrictive than necessary for many practical applications. In scenarios where the focus lies on the time evolution of a specific observable, global phases have no physical consequence. In case the global phase is of importance, its value can easily be extracted and corrected within a postprocessing step. Consequently, relaxing the requirement for the quantum circuit to determine a fixed global phase can simplify the optimization landscape.
Within our study, we make use of a less stringent cost function allowing for different global phases, namely the Hilbert-Schmidt test \cite{khatri2019quantum} given by
\begin{equation}
    \label{eq:costHS}
    \costHS(\Uref,W) = 1 - \frac{1}{d^2} \left|\overlap\right|^2.
\end{equation}
$\costHS$ is directly related to the fidelity between two states $\Uref |\psi\rangle$ and $W |\psi\rangle$ averaged over the Haar distribution~\cite{khatri2019quantum}, and hence quantifies how close the unitaries $\Uref$ and $W$ are (up to a global phase).
The two cost functions $\costF$ and $\costHS$ are related via the following inequality:
\begin{align*}
    \costHS(\Uref, W) &= 1 - \frac{1}{d^2} \left(\left[ \Re\overlap \right]^2 - \left[ \Im\overlap \right]^2\right) \nonumber\\
    &\leq 1 - \left[ 1 - \costF(\Uref,W) \right]^2 \nonumber \\
    &\leq 2\,\costF(\Uref,W).
\end{align*}
For simplicity, we denote $\cost=\costHS$, and its variational part is given by
\begin{align}\label{eq:lossHS}
    \mc{L}(\Uref, W) &= -\left|\overlap\right|^2 \\
    &= -\left[  \Re \overlap \right] ^2 -  \left[\Im \overlap \right]^2.\nonumber
\end{align}

\paragraph*{Formulation of optimization problem.}
The objective is to optimize a set of unitary two-qubit gates $G=\{G_i\}_{i\in\ind{W}}$ such that the resulting optimized brickwall circuit $W$ minimizes its distance to the reference $\Uref$. 
The underlying optimization problem can hence be expressed as
\begin{equation}
\label{eq:optimization-problem}
G_\text{opt} = \underset{G\in\mc{U}(4)^{\times N}}{\arg\min} \cost(\Uref, W).
\end{equation}
In the latter, $\mc{U}(4)$ denotes the set of unitary $4\times 4$-matrices: 
\begin{equation}
    \mc{U}(4) = \{ G_g \in \mathds{C}^{4\times 4} | G_g^\dagger G_g = \mathds{1}_4 \},
    \label{eq:unitaries}
\end{equation}
which can be viewed as a particular case of the complex Stiefel manifold. 

\paragraph*{Euclidean gradient of the cost function.}
The presented optimization problem shall be solved by means of gradient-based optimization and therefore, computing derivatives of the cost function is essential. 
While we will later see that the Euclidean gradient alone is not suitable for our Riemannian optimization method, in this part of the paper, we first present the relevant Euclidean gradient. 
A detailed calculation is given in \cref{ap:gradient-computation}. 

The gradient of $\cost$ and $\mc{L}$ are equal up to a constant factor, and hence the quantity of interest is given by $\grad \mc{L}$. 
By using the Wirtinger formalism~\cite{koor2023short}, the partial derivatives with respect to an entry of $W$ are given by
\[
    \partial _{W_{jk}} \Re \overlap
    = \frac{U^\ast_{jk}}{2} \;\;\;\;\; \text{and} \;\;\;\;\; \partial_{W_{jk}}\Im \overlap
    = \frac{U^\ast_{jk}}{2i}.
\]
The partial derivative of $\mc{L}$ with respect to $W_{jk}$ can then be computed as
\begin{align*}
    \partial_{W_{jk}}\mc{L} = -\left[\overlap U_{jk}\right]^\ast.
\end{align*}
By applying the chain rule of the Wirtinger formalism, the partial derivatives of $\mc{L}$ with respect to a specific gate are given by
\begin{align*}
    \partial_{G_i^\ell} \mc{L} = -\overlap^\ast \cdot \partial_{G_i^\ell}\overlap,
\end{align*}
where we denote the number of layers in the brickwall circuit with $L$ and the number of gates in layer $\ell$ with $n_\ell$. 
Finally, the overall gradient can be obtained as
\begin{equation}\label{eq:gradient-loss}
\grad\mc{L} = -2\,\overlap \cdot \partial_{G_i^\ell}\left[\overlap\right]_{\substack{\ell=1,\dots,L\\i=1,\dots,n_\ell}}^\ast\,.
\end{equation}
In order to compute $\grad \mc{L}$, it is necessary to evaluate both $\overlap$ and $\partial_{G_i^\ell}\overlap$. 
Conveniently, by computing $\overlap$, the cost function $\cost$ given in \cref{eq:costHS} can be directly extracted.

\paragraph*{Quantum circuit initialization.}
Previous analyses have shown that a good initialization can significantly enhance quantum circuit optimization~\cite{kotil2024riemannian,causer2024scalable,gibbs2024deep}. 
If the brickwall circuit to be optimized has $L$ layers, the initialization is chosen as follows:
\begin{itemize}[label=$\circ$]
    \item We first determine if and, if yes, which orders of (repeated) Trotterization match the number of layers $L$. If several Trotter circuits of different orders match $L$, we choose the one that yields the best approximation error. We consider Trotterizations of orders two and four. 
    \item If the brickwall circuit does not match any (repeated) Trotterization, we test whether it can be initialized as the concatenation of a second-order and a fourth-order Trotter circuit. To this end, the Trotter time steps of both Trotterizations, $t_1$ and $t_2$, have to be chosen, such that $t=t_1+t_2$ for the target simulation time $t$. We choose $t_1$ and $t_2$ such that they minimize the approximation error. 
    \item Lastly, we also allow for extending the above initialization circuits by layers consisting of identity gates to be able to initialize any brickwall circuit of arbitrary $L$. 
\end{itemize}
\section{Riemannian optimization on the complex Stiefel manifold \label{sec:riemannian}}
As described above, the goal is to optimize a set of two-qubit gates under the constraint of unitarity, which means working on the complex Stiefel manifold. One effective method for such optimization is the gradient-based Riemannian optimization, which does not necessitate any specific parameterization of the unitary gates. 
Recently, this approach has been explicitly utilized for various optimization tasks in quantum technologies~\cite{kotil2024riemannian,luchnikov2021qgopt,luchnikov2021riemannian,godinez2024riemannian,rogerson2024quantum,wiersema2023optimizing,ahmed2023gradient,melnikov2023quantum,akshay2024tensor,guo2025efficient,termanova2024tensor}. 
In the following, we will provide a brief overview of the basic concepts required for Riemannian optimization of brickwall circuits. 
For a more detailed introduction to these concepts, we refer interested readers to Refs.~\cite{hauru2021riemannian,absil2008optimization,edelman1998geometry}. 
Furthermore, we adapt the well-known ADAM optimizer~\cite{diederik2014adam} to the setting of Riemannian optimization. 

\subsection{Concepts of Riemannian geometry}
\begin{figure}
    \centering
    \includegraphics[height=4.5cm]{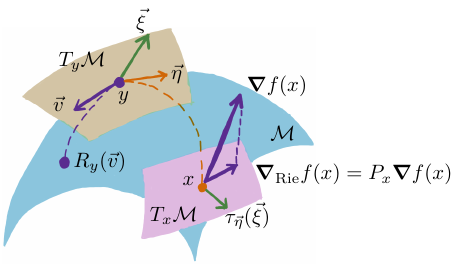}
    \caption{Visualization of concepts on the Riemannian manifold $\mc{M}$. A point $y\in\mc{M}$ is ``moved'' in the direction of the vector $\vec{v}\in T_y\mc{M}$ via the retraction $R$. A vector $\vec{\xi}$ is moved in the direction of a vector $\vec{\eta}\in T_y\mc{M}$ to the tangent space $T_x \mc{M}$ in point $x\in\mc{M}$ using the vector transport $\tau$. Lastly, the Riemannian gradient $\Riegrad f(x)$ can be obtained as the projected Euclidean gradient $\grad f(x)$ through the projector $P$.}
    \label{fig:riemannian-geometry}
\end{figure}
Intuitively, a manifold $\mc{M}$ can be thought of as a curved space that, at any point, $x\in\mc{M}$, can be locally described by a Euclidean space, which is isomorphic to a vector space called the \textit{tangent space} $T_x\mc{M}$. 
The collection of all points in the manifold and the corresponding tangent spaces, $(x, T_x\mc{M})$, is called the \textit{tangent bundle} $T\mc{M}$. 
If each tangent space is equipped with an inner product $\langle\cdot , \cdot\rangle_x: T_x\mc{M}\rightarrow T_x\mc{M}$, $\mc{M}$ and $\langle\cdot , \cdot\rangle$ form a \textit{Riemannian manifold}. 
When naively moving a point $x\in\mc{M}$ along the direction of a vector $\vec{v}_x\in T_x\mc{M}$, the resulting point will, in general, lie in $T_x\mathcal{M}$ and not be an element of $\mc{M}$ anymore. 
A \textit{retraction}
\[
R:T\mc{M}\rightarrow \mathcal{M},\; R_x(\vec{v}_x)=y
\] 
is a smooth map from the tangent bundle into the manifold, and hence $R_x(v_x)$ corresponds to moving a point $x$ in the direction of $v_x$ within the manifold. 
Moreover, a so-called \textit{vector transport} 
\[\tau_{\vec{\eta}_x} : T_x\mc{M} \rightarrow T_y\mc{M},\; \tau_{\vec{\eta}_x}(\vec{v}_x)=\vec{v}_y\] 
is a smooth map that moves a vector $\vec{v}_x \in T_x\mc{M}$ to another tangent space $T_y\mc{M}$ in point $y\in\mathcal{M}$, where $\vec{\eta}_x\in T_x\mc{M}$ is the direction of transport. 
Consider a Riemannian manifold $\mc{M}$ and a function $f:\mc{M}^{\times n}\rightarrow\mathds{R}$ with inputs $x=\{x_i\}_{i=1}^n$ from a product manifold. 
The so-called \textit{Riemannian gradient} can be obtained from the Euclidean gradient $\grad f(x)$ by projecting the latter entry-wise onto $T_x\mc{M}^{\times n}$ via some suitable projection $P$: 
\[
\Riegrad f(x) = P_x \grad f(x) \in T_x\mc{M}^{\times n}.
\]
An overview of these concepts is visualized in \cref{fig:riemannian-geometry}.

\subsection{Example of the complex Stiefel manifold}
As already mentioned, $\mathcal{U}$ defined in \cref{eq:unitaries} is a manifold, and the tangent space in a point $G\in\mathcal{U}$ can be parameterized by a set of complex anti-Hermitian matrices
\[
    T_G\,\mathcal{U} = \{ GA: A\in\mathds{C}^{m\times m}, A^\dagger = -A \}.
\]
Furthermore, a Riemannian metric can be defined on each tangent space $T_G\mathcal{U}$ by
\begin{equation}\label{eq:inner-product}
    \langle \cdot,\cdot\rangle : T_G\,\mathcal{U}\times T_G\,\mathcal{U}\rightarrow \mathds{R}, \;\;\;\langle X,Y\rangle_G = \text{Tr}(X^\dagger Y),
\end{equation}
with which $\mathcal{U}$ can be identified as a Riemannian manifold and -- more specifically -- a particular case of the complex Stiefel manifold. 
Note that the cost function defined in \cref{sec:setup} involves $\overlap=\langle \Uref, W\rangle$. 
A proper \textit{retraction of $\mathcal{U}$} is induced by the polar decomposition $A=UM$ with $U$ being a unitary matrix and $M$ being a positive semi-definite Hermitian matrix. 
When denoting the corresponding unitary part as $\pi(A)=U$, a vector $\vec{\xi}\in T_G\,\mathcal{U}$ can be retracted into the unitary manifold via
\begin{equation}
    R^\text{polar}:T\mathcal{U}\rightarrow \mathcal{U}, \;\;\; R_G^\text{polar}(\vec{\xi}) = \pi(G+\vec{\xi}).
    \label{eq:polar-retraction}
\end{equation}
A general vector $\vec{\eta}\in\mathds{C}^{m\times m}$ can be \textit{projected onto the tangent space $T_G\,\mathcal{U}$} by means of the linear projector
\begin{align}
    P_G(\vec{\eta}) &= \vec{\eta} - \frac{1}{2} G (\vec{\eta}^\dagger G + G^\dagger \vec{\eta}) \nonumber \\
    &= G\, \text{skew}(G^\dagger \vec{\eta}).
    \label{eq:projector}
\end{align}
Utilizing \cref{eq:projector}, the map
\begin{equation}
    \tau_{\vec{\eta}}(\vec{\xi}) = P_{R_G^\text{polar}(\vec{\eta})}(\vec{\xi})
\end{equation}
is a valid \textit{vector transport on $\mathcal{U}$}. 
In other words, a vector $\vec{\xi}\in T_G\,\mathcal{M}$ in the tangent space corresponding to $G\in\mathcal{M}$ is transported in the direction of $\vec{\eta}\in T_G\,\mathcal{M}$ by projecting it onto the tangent space corresponding to the retracted directional vector $R_G^\text{polar}(\vec{\eta}) \in \mathcal{U}$ using \cref{eq:projector}. 

When optimizing a brickwall circuit $W$ consisting of $|W|$ two-qubit gates $\{G_i\}_{i\in\ind{W}}$, the cost function $f$ is defined on the product manifold $\mc{U}^{\times |W|}$
\[
f: \mc{U}^{\times |W|} \rightarrow \mathds{R}.
\]
The corresponding tangent space is given by the direct sum of the individual tangent spaces 
\[
T_W\mc{U}^{\times |W|} = \bigoplus_{i\in\ind{W}}T_{G_i}(\mc{U}).
\]
Using the same projector as given in \cref{eq:projector} the corresponding \textit{Riemannian gradient} can be computed via
\begin{equation}
    \Riegrad f(W) = P_W \grad f(W). \label{eq:Riegrad}
\end{equation}
Lastly, the overall retraction corresponds to applying the retraction in \cref{eq:polar-retraction} to the individual quantum gates and tangent vectors. 

\subsection{Adapting the ADAM optimizer to the Riemannian setting of the complex Stiefel manifold \label{subsec:RieADAM}}
In the following, we make use of the previously introduced concepts to adapt the first-order gradient-based ADAM optimizer~\cite{diederik2014adam} to the Riemannian optimization setting on the manifold of unitary operations~\cite{becigneul2018riemannian}. 
ADAM aims to minimize an objective function $f$ via the component-wise update rule
\[
    x_{t+1} \leftarrow x_t - \alpha m_t / \sqrt{v_t},
\]
where $\alpha\in\mathds{R}$ is the learning rate, and $m_t$ and $v_t$ are the first and second momenta in each iteration $t$, which can be obtained from the gradient $\grad f$ as 
 \[
    m_t = \beta_1m_{t-1} + (1-\beta_1)\grad f,
\]
\[
    v_t = \beta_2 v_{t-1} + (1-\beta_2)(\grad f)^2,
\]
where usually $\beta_1=0.9$ and $\beta_2=0.99$ are chosen. 

\paragraph*{Using the Riemannian gradient.}
To constrain ADAM to the unitary manifold, firstly, $\grad f$ has to be replaced by $\Riegrad f$, from which follows that $v_t$ is computed from the inner product $\langle\Riegrad f, \Riegrad f\rangle_{x_t}$. 
Additionally, some further modifications need to be made. 

\paragraph*{Vector transport of momentum.}
$m_{t}$ is computed by adding the scaled Riemannian gradient $\Riegrad f \in T_{x_t}\mc{M}$ to the scaled previous momentum $m_{t-1}\in T_{x_{t-1}}\mc{U}$, i.e., two quantities from different tangent spaces. 
Therefore, it is necessary to transport $m_{t-1}$ to the tangent space of the current point. 
The modified Riemannian momenta updates are then given by
\[
    \Tilde{m}_t = \beta_1 \tau_{\Tilde{m}_t}(\Tilde{m}_{t-1}) + (1-\beta_1) \nabla_\text{Rie}f,
\]
\[
    \Tilde{v}_t = \beta_2 \Tilde{v}_{t-1} + (1-\beta_2) \langle\Riegrad f, \Riegrad f\rangle_{x_t}.
\]

\paragraph*{Retraction of updated step.}
Since the Riemannian momenta determine the update direction, the updated point $x_{t+1}$ will lie in the tangent space of the former point $x_t$. 
Therefore, the retraction given in \cref{eq:polar-retraction} must be applied to obtain valid new parameters within $\mathcal{M}$. 
The modified update rule is hence given by
\[
    \Tilde{x}_{t+1} \leftarrow R_{\Tilde{x_t}}^\text{polar}\left(\Tilde{x}_t - \alpha \Tilde{m}_t / \sqrt{\Tilde{v}_t}\right) \in \mathcal{M}.
\]
Note that the ADAM optimizer and its Riemannian adaption are invariant under gradient rescaling.

\section{Tensor networks methods for quantum circuit optimization\label{sec:tn-methods}}

\subsection{Merging a brickwall layer of two-qubit gates into a matrix product operator \label{subsec:merging-mpo-with-layer}}
\begin{figure*}
  \centering
  \begin{subfigure}[b]{0.9\textwidth}
    \includegraphics[width=\textwidth]{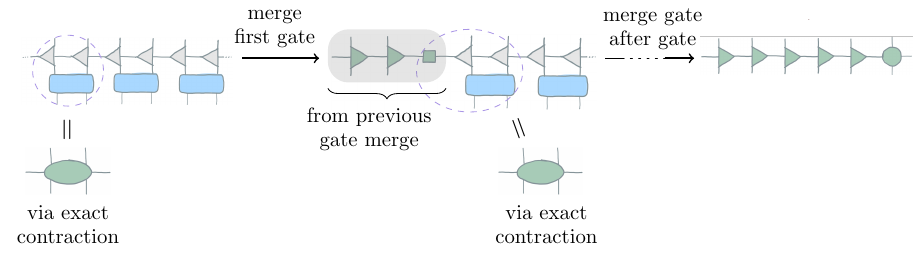}
    \caption{A brickwall layer can be merged into a right-canonicalized \ac{MPO} via a left-to-right-sweep The canonical center of the \ac{MPO} is shifted accordingly during the left-to-right-sweep. The final \ac{MPO} is right canonicalized.}
  \end{subfigure}
  \hfill
  \begin{subfigure}[b]{0.85\textwidth}
    \includegraphics[width=0.65\textwidth]{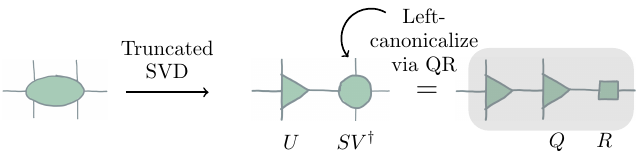}
    \caption{How to merge a two-qubit gate into the corresponding local \ac{MPO} tensors.}
  \end{subfigure}
  \caption{Illustration of how to merge a brickwall layer into an \ac{MPO} in a left-to-right-sweep using the tensor network framework. Similarly, a right-to-left-sweep can be used, which is not explicitly illustrated here.}
  \label{fig:merge-mpo-with-layer}
\end{figure*}
A fundamental computational step in previous and the presented methods involves merging a brickwall layer of two-qubit gates into an \ac{MPO}. 
This operation arises in key contexts such as approximating the time-evolution operator $U_t$ as an MPO and computing both the cost function and its gradient. Given its central role, we now describe this procedure in more detail. 

Recall that any \ac{MPO} can be brought into left- or right-canonical form via a sweep of QR or RQ decompositions, respectively -- performed from left to right or right to left across the \ac{MPO} tensors, i.e.,
\begin{equation*}
    \includegraphics[width=0.9\linewidth]{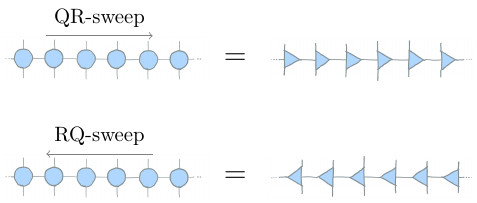},
\end{equation*}
where \raisebox{-0.2em}{\includegraphics[height=1.2em, rotate=90]{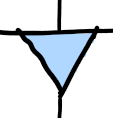}}\; denotes a left-isometry and \raisebox{0.9em}{\includegraphics[height=1.2em, rotate=-90]{gfx/canonical-tensor.png}} \; denotes a right-isometry. 

The objective is to merge a brickwall layer of two-qubit gates into an \ac{MPO} while simultaneously truncating the bond dimension to a maximum of $\maxbondim$. 
The procedure either begins with a right- or left-canonicalized \ac{MPO}, where the canonical center is located at the leftmost or rightmost tensor, respectively. 
For simplicity, we will only describe the method for an initial right-canonicalized \ac{MPO}. 
The two-qubit gates are then sequentially merged from left to right as outlined by the following steps:
\begin{itemize}
    \item The considered two-qubit gate is exactly contracted with the corresponding pair of local \ac{MPO} tensors and some residual tensor $R$ obtained from the previous quantum gate merge. If the considered quantum gate corresponds to the leftmost gate, $R$ is ignored. This yields an intermediate higher-rank tensor \( K \).
    
    \item The tensor \( K \) is then decomposed via a singular value decomposition (SVD), \( K = USV^\dagger \). To optimally truncate the bond dimension, only the largest \(\maxbondim\) singular values are retained. The updated left tensor is set to \( K_1 = U \), which is left-isometric by construction, while the updated right tensor is given by \( K_2 = \tilde{S}V^\dagger \), where \( \tilde{S} \) denotes the truncated diagonal matrix of singular values.
    
    \item To restore a mixed-canonical form, the left-isometric component of \( K_2 \) is extracted via a QR decomposition, \( K_2 = QR \), where \( Q \) is the new left-canonical tensor.
    
    \item The residual tensor \( R \) is then contracted with the next two-qubit gate and the subsequent pair of local \ac{MPO} tensors. This effectively shifts the canonical center to the next site pair to be merged.
\end{itemize}

This procedure is applied sequentially to each two-qubit gate in the brickwall layer. 
After completing a full left-to-right sweep, all gates have been incorporated into the \ac{MPO}, resulting in an updated MPO in left-canonical form. 
The subsequent brickwall layer can then be merged analogously via a right-to-left sweep. 
An illustration of the left-to-right merging process is provided in \cref{fig:merge-mpo-with-layer}.

\subsection{Matrix product operator representation of the reference time evolution operator\label{subsec:MPO-reference}}

To solve the optimization problem formulated in \cref{eq:optimization-problem}, a reference operator representing the time evolution propagator $U_t=e^{-iHt}$ is required. 
While $U_t$ can be computed for small systems via numerically exact diagonalization, this approach is infeasible for larger systems of interest due to the curse of dimensionality. 
Therefore, we choose to approximate $U_t$ to a reasonable accuracy and efficiently express it as an \ac{MPO} with a maximum bond dimension $\maxbondim$~\cite{causer2024scalable,gibbs2024deep}. 

\begin{figure}
    \centering
    \includegraphics[width=0.95\linewidth]{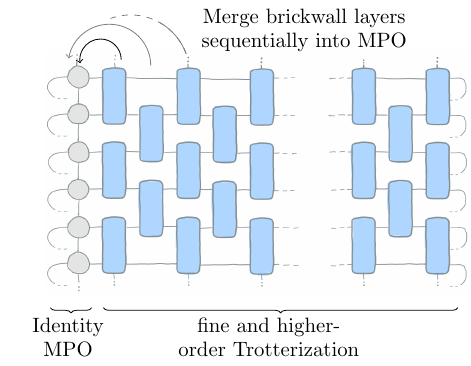}
    \caption{How to obtain the reference \ac{MPO} via a suitable fine higher-order Trotterization for system sizes of $N\geq 12$. The corresponding deep Trotter circuit is interpreted as a tensor network, and its brickwall layers are sequentially merged into an initial identity \ac{MPO}.}
    \label{fig:reference-computation}
\end{figure}

\paragraph*{Computing the reference matrix product operator.}
For the optimization, it is essential to have an efficient and accurate representation of the time evolution operator \( U_t = e^{-iHt} \) as a reference, which we will denote as \( \Uref \). Depending on the number of qubits \( N \), the references are generated in two different ways:
\begin{itemize}[label=$\circ$]
    \item For system sizes of $N\leq 12$, we can compute $U_t$ numerically exactly as a full-rank matrix and decompose and truncate it to an \ac{MPO}. 
    \item For larger systems, we obtain an accurate approximation of $U_t$ using $n$ repetitions of a fourth-order Trotterization $\Trotter{IV}{,n}$. For this purpose, $n$ is chosen such that $\Trotter{IV}{,n}$ has a negligible approximation error for the systems considered in this work. We then treat the resulting deep quantum circuit as a tensor network and contract it into an \ac{MPO}~\cite{causer2024scalable,gibbs2024deep}. To this end, we start with an initial identity \ac{MPO} and sequentially merge the layers of $\Trotter{IV}{,n}$ into it as explained in \cref{fig:reference-computation}. The procedure of merging a brickwall layer into an \ac{MPO} follows the same strategy discussed in \cref{subsec:merging-mpo-with-layer}.
\end{itemize}

In both cases, the bond dimensions of the \ac{MPO} are truncated to some maximum value $\maxbondim$~\cite{schollwock2011density}, and we denote the resulting \ac{MPO} by $U_{\maxbondim}$ to indicate this property. 
In practice, the accuracy of $\Uref$ does not have to be arbitrarily high. 
Suppose the final error between the optimized quantum circuit and the reference is $\errorFinal$. 
In that case, it is sufficient to have a reference operator $\Uref$ that approximates the exact time evolution $U_t$ by 
\[
\cost(U_t,\Uref)\approx\frac{\errorFinal}{10}=\errorThres.
\]
To further relax computational demands, a reasonable target error $\errorFinal$ is evaluated, and $U_{\maxbondim}$ is compressed to the smallest bond dimension $\chi\leq\maxbondim$~\cite{gibbs2024deep} for which it still holds that 
\[
\cost(U_{\maxbondim},U_\chi) \leq\errorThres.
\]
The compression is achieved by applying a \ac{SVD} sweep on $U_{\maxbondim}$ in mixed-canonical form, where the canonical center is always moved to the local tensor that is currently truncated~\cite{schollwock2011density}.

\paragraph*{Quantifying the reference approximation error.}
 Several error sources lead to a deviation of $\Uref$ from $U_t$. 
\begin{itemize}[label=$\circ$]
    \item Trotterization error $\errorTrotter$ when approximating $U_t$ by $\Trotter{IV}{,n}$ for systems with $N>12$ qubits. 
    \item Truncation error $\errorTruncation$ when computing $U_{\maxbondim}$.
    \item Compression error $\errorCompression$ when compressing $U_{\maxbondim}$ to $U_\chi$, which is given by the chosen $\epsilon_\text{thres}$.
\end{itemize}
To estimate the approximation error $\errorTrotter$, we note that for local Hamiltonians, the error is expected to scale linearly in system size. 
Hence, $\errorTrotter$ for large system sizes can be estimated from extrapolating the approximation errors of a Trotterization for a considered simulation time $t$ for small enough systems, for which the exact full-rank matrix $U_t$ can still be computed. 
In order to quantify the truncation error $\errorTruncation$ of $U_{\chi_\text{max}}$, a series of \acp{MPO} $\{U_\chi\}_{\chi > \chi_\text{max}}$ is computed as described before. 
First, it is verified that increasing the bond dimension does not change the \ac{MPO} significantly:
\[
\cost(U_\chi, U_{\chi+1}) < 10^{-10}.
\]
The $\errorTruncation$ of $U_{\chi_\text{max}}$ is then given by $\cost(U_{\chi_\text{max}}, U_{\chi_\text{max}+1})$. 
Lastly, the compression error $\errorCompression$ is determined by the choice of $\epsilon_\text{thres}$. 
This study compares the optimization results with Trotterizations up to the fourth order. 
Hence, $\errorCompression$ is chosen such that the reference \ac{MPO} is accurate enough to reproduce the error scaling behavior of these Trotter circuits. 
We will accordingly document the error estimations of the $\Uref$ used in the numerical simulations.

\subsection{Tensor network methods for cost function and gradient evaluation}\label{subsec:tn-implementation}
In the following, we present the tensor network methods used to evaluate both the gradient and the cost function presented in \cref{sec:setup}. 

\paragraph*{Computing the cost function in the context of tensor networks.}
\begin{figure*}[htb]
    \centering
    \includegraphics[width=0.95\textwidth]{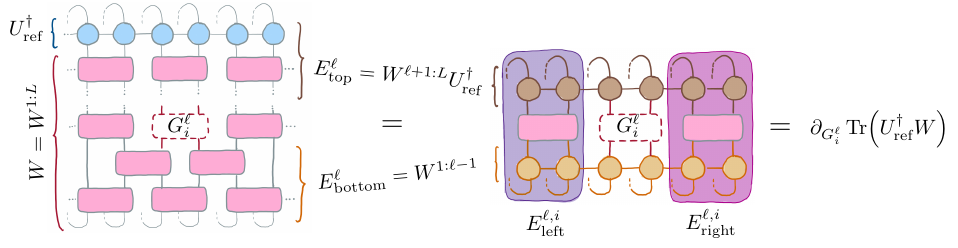}
    \caption{Diagrammatic sketch of $\partial_{G_i^\ell} \Tr(\Uref^\dagger W)$, where $G_i^\ell$ is the $i$-th gate in the $\ell$-th layer of the brickwall circuit. The partial derivative can be obtained by ``cutting out'' the considered gate $G_i^\ell$. For the efficient evaluation, a top environment \ac{MPO} $E_\text{top}^{\ell}$ and a bottom environment \ac{MPO} $E_\text{bottom}^\ell$ are computed. From there, the left and right environment, $E_\text{left}^{\ell,i}$ and $E_\text{right}^{\ell,i}$, are evaluated. Finally, contracting the resulting tensor network can compute the partial derivative.}
    \label{fig:gradient-computation}
\end{figure*}
When interpreting the brickwall circuit as a tensor network, the partial derivatives $\partial_{G_i^\ell}\overlap$ can be obtained by ``cutting out'' the respective quantum gate $G_i^\ell$ from \cref{eq:overlap}. 
This is visualized in \cref{fig:gradient-computation}. 
The cost function is given by \cref{eq:costHS} with the overlap $\overlap$ being representable by a tensor network contraction as given in \cref{eq:overlap}. 
Given any partial derivative $\partial_{G_i^\ell} \overlap$, $\overlap$ can quickly be evaluated by contracting $\partial_{G_i^\ell} \overlap$ with the corresponding ``cut-out-gate'' $G_i^\ell$. 

\paragraph*{Computing the partial derivatives: Identifying tensor networks environments.}
To compute $\partial_{G_i^\ell}\overlap$, we follow an approach similar to Ref.~\cite{gibbs2024deep}. 
For the partial derivative $\partial_{G_i^\ell}\overlap$ with respect to the $i$-th quantum gate in layer $\ell$, we first compute a top environment \ac{MPO} $\topenv^\ell$, which is obtained by merging all layers $\ell'=\ell+1,\dots, L$ lying above the considered layer $\ell$ into the adjoint reference \ac{MPO} $\Uref^\dagger$. 
Similarly, a bottom environment \ac{MPO} $\bottomenv^\ell$ is computed by merging all layers $\ell'=1,\dots, \ell-1$ below the considered layer $\ell$. 
This leads to our layer $\ell$ being ``sandwiched'' by the two \acp{MPO} $\topenv^l$ and $\bottomenv^\ell$ as visualized in \cref{fig:gradient-computation}. 
To contract this resulting tensor network, two further environments are computed. 
The first results from contracting all tensors to the left of the considered gate $G_i^\ell$, which we call the left environment $\leftenv^{\ell,i}$. 
Similarly, the right environment $\rightenv^{\ell,i}$ is obtained by contracting all tensors to the right of $G_i^\ell$. 
\paragraph*{Computing the partial derivatives: Environment caching.}
For the efficient evaluation of all $\partial_{G_i^\ell}\overlap$, it is helpful to cache some of the environments. 
Note that for a fixed layer $\ell$, all left environments $\leftenv^{\ell,i}$ (right environments $\rightenv^{\ell,i}$) can be recursively obtained by a left-to-right (right-to-left) sweep, since
\begin{equation*}\label{eq:left-env-recursion}
    \includegraphics[height=3cm]{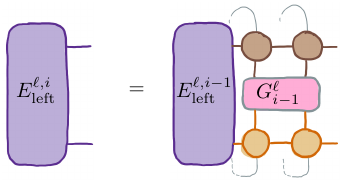}
\end{equation*}
and
\begin{equation*}\label{eq:right-env-recursion}
	\includegraphics[height=3cm]{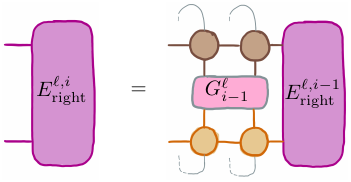}.
\end{equation*}
Using efficient tensor contractions, each evaluation of $E_\text{left}^{\ell,i}$ ($E_\text{right}^{\ell,i}$) has a cost that scales the most with $\mc{O}\left( D^3 \chi^3 \right)$, where in our case it is always $D=2$. Since each two-qubit gate has to be merged once in order to compute all $E_\text{left}^{\ell,i}$ ($E_\text{right}^{\ell,i}$), the overall cost per gradient computation scales linearly in the number of gates $|W|$, i.e., $\mc{O}\left( |W| d^3 \chi^3 \right)$.

Furthermore, $\topenv^\ell$ ($\bottomenv^\ell$) for all layers $\ell=1,\dots,L$ in the brickwall circuit can be recursively computed by a top-to-bottom (bottom-to-top) sweep, since
\begin{gather*} \label{eq:top-env-recursion}
    E_\text{top}^{\ell-1}\;\;\;\; = E_\text{top}^\ell W^\ell \\
    \Leftrightarrow \\
    \begin{minipage}{0.49\textwidth}
	\vspace{-8pt}
    \centering
    \includegraphics[height=1.5cm]{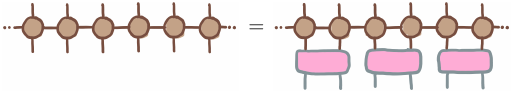},
   \end{minipage}	
\end{gather*} 
and
\begin{gather*}\label{eq:bottom-env-recursion}
    E_\text{bottom}^{\ell+1} = E_\text{bottom}^\ell W^\ell \\
    \Leftrightarrow\;\;\;\;\; \\
    \begin{minipage}[h]{0.49\textwidth}
	\vspace{-8pt}
	\includegraphics[height=1.5cm]{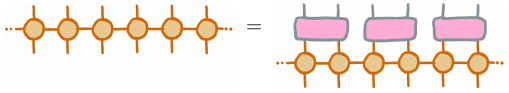},
   \end{minipage}	
\end{gather*}
with the initial environments $\topenv^L=\Uref^\dagger$ and $\bottomenv^1=\text{id}_\text{MPO}$. 
To this end, a brickwall layer has to be merged into the previous environment \ac{MPO}, which is achieved by the strategy discussed in \cref{subsec:merging-mpo-with-layer}.
The contraction cost to merge a two-qubit gate with the corresponding local \ac{MPO} tensors scales at most with $\mc{O}\left( D^5 \chi^3 \right)$. The resulting merged tensor needs to be split using a truncated \ac{SVD} in order to retain the \ac{MPO} form and a restricted bond dimension. Such an \ac{SVD} contributes a cost of $\mc{O}\left( [D^2 \chi]^3 \right)$. 
Since each gate has to be merged and split once into an \ac{MPO} in order to compute all $E_\text{bottom}^{\ell}$ ($E_\text{top}^{\ell}$), the overall cost within each gradient evaluation scales linearly in the number of gates $|W|$, i.e., $\mc{O}\left( |W| D^6 \chi^3 \right)$.

\paragraph*{Full computation of the gradient $\grad\loss$.}
In practice, the computation of $\grad \mc{L}$ consists of
\begin{enumerate}[label=(\roman*)]
    \item Compute all partial derivatives $\partial_{G_i^\ell}\overlap$.
    \item Compute the overlap $\overlap$ from one of the partial derivatives.
    \item Compute $\grad\loss$ via \cref{eq:gradient-loss}.
\end{enumerate}

Computing all $\partial_{G_i^\ell}\overlap$ in step (i) can be achieved by the following strategy: 
\begin{enumerate}
    \item Perform a top-to-bottom sweep to compute and store all $\topenv^\ell$.
    \item Perform a bottom-to-top sweep and consecutively compute all partial derivatives with respect to the gates in the layers $\ell=1,\dots,L$.
    \begin{enumerate}
        \item At each layer $\ell$, first compute $\bottomenv^\ell$ from the previous $\bottomenv^{\ell-1}$. Only the current and previous bottom environments must be stored during this sweep.
        \item Perform a right-to-left sweep to compute and cache $\rightenv^{\ell,i}$ for all $i=n_\ell,\dots,1$.
        \item Perform a left-to-right-sweep to compute each $\partial_{G_i^{\ell}}\overlap$ by first computing $\leftenv^{\ell,i}$ from the previous $\leftenv^{\ell,i-1}$ and then contracting the resulting tensor network. Only the previous and current left environments have to be stored in this sweep.
    \end{enumerate}    
\end{enumerate}

The last contraction in step~(c) has a cost that scales at most with $\mc{O}\left( D^3 \chi^3 \right)$. Since this contraction has to be done for each two-qubit gate, the overall cost to evaluate the full gradient coming from this step scales with $\mc{O}\left( |W| D^3\chi^3 \right)$. Taking into account the intermediate contractions to evaluate all environments, the cost to compute the gradient scales at most with $\mc{O}\left( |W| D^6 \chi^3 \right)$.
An overview of the contraction costs per gradient evaluation is given in \cref{tab:contraction-cost}.

\begin{table}[h]
\caption{\label{tab:contraction-cost}
Overview of contributing contraction costs per gradient evaluation with respect to the number of two-qubit gates $|W|$, the maximum bond dimension $\chi$ and the qubit dimension $D=2$.
}
\centering
\begin{tabular}{ll}
\hline\hline
$E_\text{top}$/$E_\text{bottom}$ & $\mathcal{O}\left( |W| D^6 \chi^3 \right)$\\
$E_\text{left}$/$E_\text{right}$ & $\mathcal{O}\left( |W| D^3 \chi^3 \right)$\\
Final contraction & $\mathcal{O}\left( |W| D^3 \chi^3 \right)$\\
\hline\hline
\end{tabular}
\end{table}

The procedure of computing one partial derivative is visualized for a specific gate $G_i^\ell$ in \cref{fig:gradient-computation}. 
The Riemannian gradient $\Riegrad \mc{L}$ is then obtained through the projection given in \cref{eq:projector} as discussed in \cref{sec:riemannian}.

\subsection{Enabling gradient-based parameter optimization\label{subsec:euclidean-optimization}}
As discussed in the introduction, an alternative strategy for quantum circuit optimization involves parameterizing quantum gates~\cite{mansuroglu2023variational,mansuroglu2023problem,tepaske2023optimal,mc2023classically,mc2024towards,robertson2023approximate}. While a geometric-aware Riemannian optimization approach is expected to perform better especially on difficult loss landscapes. parameterization-based techniques are particularly effective when adapting to native hardware gate sets. 
For instance, certain quantum hardware platforms natively support two-qubit gates of the form
\[
V(a,b,c) = e^{i(aXX + bYY + cZZ)}.
\]
In such cases, it is advantageous to directly optimize the Weyl decomposition~\cite{kraus2001optimal}:
\[
G = (K_1 \otimes K_2) \cdot V(a,b,c) \cdot (K_3 \otimes K_4),
\]
where each $K_i \in SU(2)$ and can be expressed as a sequence of single-qubit rotations, e.g.,
\[
K_i = RZ(\theta_i)\, RY(\psi_i)\, RZ(\phi_i).
\]
This yields a 15-parameter representation of each two-qubit gate: 
\[
\{\alpha_i\}_{i=1}^{15} = \{a, b, c\} \cup \{\theta_i, \psi_i, \phi_i\}_{i=1}^4.
\]

Our scalable tensor network-based method for the evaluation of \cref{eq:gradient-loss} presented in \cref{subsec:tn-implementation} can readily be used in the parameterized context as well. 
For a gradient-based Euclidean optimization, the gradient with respect to the parameters $\{\alpha_i\}$ is required. 
The latter can be obtained by a simple chain rule:
\begin{equation}\label{eq:gradient-param}
    \frac{\partial \mathcal{L}}{\partial\alpha_i} = \left\langle\frac{\partial L}{\partial G}, \frac{\partial G}{\partial \alpha_i}\right\rangle,
\end{equation}
which intuitively corresponds to replacing the ``cut-out-gate'' in the tensor network picture with the corresponding inner derivative $\frac{\partial G}{\partial \alpha_i}$. 
For completeness, we have listed $\frac{\partial G}{\partial \alpha_i}$ for the case of the Weyl decomposition in \cref{ap:weyl-derivatives}. 

\section{Numerical simulations\label{sec:results}}
In the following, we consider Hamiltonians of (i) spin chains, and (ii) fermionic systems. 
We use the common notation of $(\sigma_i^1, \sigma_i^2, \sigma_i^3)=(X_i,Y_i,Z_i)$ for the vector of Pauli operators acting on the $i$-th qubit. 
Moreover, we denote in the context of second quantization the creation (annihilation) operator acting on orbital $p$ as $a_p^\dagger$ ($a_p$) and the corresponding number operator as $n_p = a_p^\dagger a_p$. 

While our method can generally be applied to non-translationally invariant systems, we mainly consider systems with translational invariance for simplicity. 
However, we explicitly demonstrate this aspect for 10 disordered transverse-field Ising models on a one-dimensional chain with 20 sites. 
Furthermore, we show the optimization results for non-disordered transverse-field Ising and Heisenberg models on a one-dimensional chain with 50 sites although our method is also applicable to larger systems. 
Regarding the fermionic systems, we first show the optimization results for a non-disordered spinful Fermi-Hubbard model on a one-dimensional chain for 50 spin orbitals. 
As a proof of concept, we then demonstrate that our method can be applied to molecular Hamiltonians using the exemplary system of \ac{LiH}. 
The utilized error measure $\cost$ is given by \cref{eq:costHS}.

\subsection{Spin systems\label{subsec:spin-systems}}
\subsubsection{Transverse-field Ising model (non-disordered).}

As a first example, we consider the transverse-field Ising model Hamiltonian on a chain with $N=50$ sites and open boundary conditions: 
\[
H^\text{Ising} = \sum_{i=1}^{N-1}  J Z_i Z_{i+1} + \sum_{i=1}^{N} \left( gX_i + hZ_i \right),
\]
with $J,g,h\in\mathds{R}$. 
In this study, we set without loss of generality $J=1$ and $g=0.75$, specifically considering the non-integrable case $h=0.6$. 
We compute $\Uref$ as 20 repetitions of a fourth-order Trotterization, $\Trotter{IV}{,20}$, for a simulation time of $t=2$, which is further compressed to a threshold of $\epsilon_\text{thres}$. 
An overview of the reference error estimation is given in \cref{ap:reference-error} \cref{tab:reference-overview}. 

The two-qubit Trotterization gates are given by 
\[
G^\text{Ising} = e^{-i\Delta t
\left(JZZ + \frac{g_1}{2} XI + \frac{g_2}{2}IX + \frac{h_1}{2}ZI + \frac{h_2}{2}IZ \right)},
\]
where 
\begin{itemize}[label=$\circ$]
    \item $g=g_1=g_2$ and $h=h_1=h_2$ for gates in the middle of the circuit,
    \item $g_2=g=\frac{g_1}{2}$ and $h_2=h=\frac{h_1}{2}$ for gates that act on the first qubit, and
    \item $g_1=g=\frac{g_2}{2}$ and $h_1=h=\frac{h_2}{2}$ for gates that act on the last qubit
\end{itemize}
 due to the open boundary conditions. 

  \begin{figure}[H]
    \centering
    \includegraphics[width=0.49\textwidth]{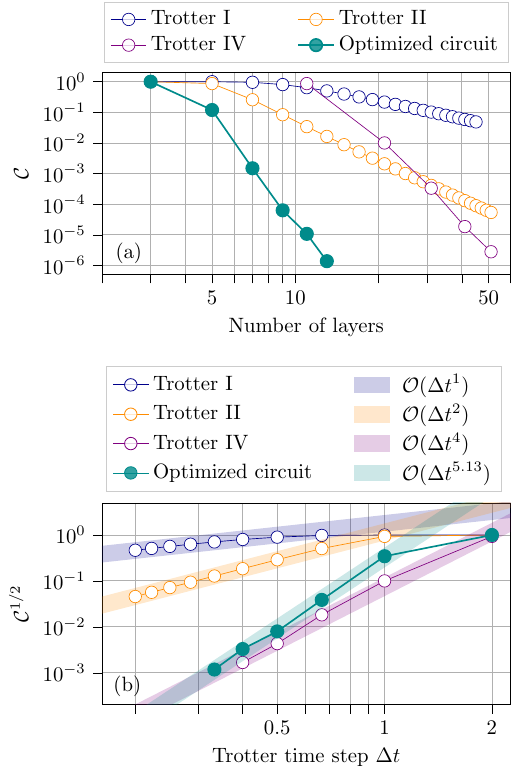}
    \caption{Results for the Ising model on a chain with $N=50$ sites, $J=1$, $g=0.75$, $h=0.6$, and $t=2$. (a)~The error of the Riemannian optimized brickwall circuit compared to that of different Trotter circuits. (b)~The error scaling in Trotter time step $\Delta t$ for the Riemannian optimized brickwall circuit compared to that of various Trotter circuits.}
    \label{fig:50_ising-results-non-disordered}
\end{figure}

The quantum circuits that are optimized are initialized as Trotterizations of order two. 
The optimization results and their comparison to Trotterizations of various orders are shown in \cref{fig:50_ising-results-non-disordered}.

The Riemannian optimization method can increase the accuracy of each optimized quantum circuit with a relative error improvement, $\cost_\text{rel}=\cost_\text{init}/\cost_\text{opt}$, of up to four orders of magnitude as shown in \cref{fig:50_ising-results-non-disordered}(a). 
For example, the brickwall circuit with 12 layers could be optimized to an accuracy of five repetitions of a fourth-order Trotterization, which would initially require 51 layers. 
Furthermore, since all considered initial quantum circuits correspond to a Trotterization of order two, we can investigate the error scaling behavior with respect to the discretized Trotter time step $\Delta t$. 
For naive Trotterization of $k$-th order, the Trotter error scales like $\mc{O}(\Delta t^k)$. 
For the optimized brickwall circuit, we observe a scaling behavior of $\mc{O}\left(\Delta t ^{5.13}\right)$, i.e., a better scaling behavior for $\Delta t \ll 1$ than a fourth-order Trotterization. 
This analysis is shown in \cref{fig:50_ising-results-non-disordered}(b).

\subsubsection{Transverse-field Ising model (disordered).}
\begin{figure}
    \centering
    \includegraphics[width=0.49\textwidth]{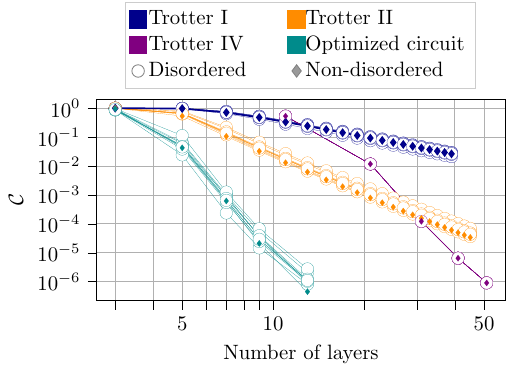}
    \caption{Results for the disordered Ising model on a one-dimensional chain of $N=20$ sites. The corresponding non-disordered Ising model is plotted for reference for parameters $J=1$, $g=0.75$, $g=0.6$, and $t=2$. The parameters for the disordered models are uniformly sampled as $J_i\in [\frac{1}{2}J, \frac{3}{2}J]$, $g_i\in [\frac{1}{2}g, \frac{3}{2}g]$, and $h_i\in [\frac{1}{2}h,\frac{3}{2}h]$.}
    \label{fig:20_ising-results-disordered}
\end{figure}

Unlike previous Riemannian quantum circuit optimization methods~\cite{kotil2024riemannian}, our new method can also be applied to systems without translational invariance. 
To demonstrate this aspect, we consider a disordered modification of the prior Ising model: 
\[
H^\text{Ising,dis} = \sum_{i=1}^{N-1}  J_i Z_i Z_{i+1} + \sum_{i=1}^N \left( g_i X_i + h_i Z_i\right)
\]
with $J_i, g_i, h_i \in \mathds{R}$, where we uniformly sample $J_i\in [\frac{J}{2}, \frac{3J}{2}]$, $g_i\in [\frac{g}{2}, \frac{3g}{2}]$, $h_i\in [\frac{h}{2}, \frac{3h}{2}]$ and without loss of generality $J,g,h$ are chosen as in the translationally invariant case. 
We generate 10 different disordered Ising models of $N=20$ sites and optimize each. 
The results for this setting are visualized in \cref{fig:20_ising-results-disordered}, where the optimization result for the corresponding non-disordered system is additionally displayed for comparison. 
Our method can handle disordered systems and achieves optimization results similar to those in the previous non-disordered case. 

\subsubsection{Heisenberg model.}
As a second spin system, we consider the Heisenberg Hamiltonian on a chain of $N=50$ sites: 
\[
H^\text{Heis} = \sum_{i=1}^{N-1} \sum_{\alpha=1}^3 J^\alpha \sigma_i^\alpha \sigma_{i+1}^\alpha + \sum_{i=1}^N \sum_{\alpha=1}^3 h^\alpha \sigma_i^\alpha,
\]
with $\vec{J},\vec{h}\in \mathds{R}^3$. 
In the numerical simulations, we consider the parameters $\vec{J}=(1,1,-\frac{1}{2})$ and $\vec{h}=(\frac{3}{4},0,0)$. 
We compute $\Uref$ as 10 repetitions of a fourth-order Trotterization, $\Trotter{IV}{,10}$, for a simulation time of $t=0.5$, which is further compressed to a threshold of $\epsilon_\text{thres}$. 
An overview of the reference error estimation is given in \cref{ap:reference-error} \cref{tab:reference-overview}. 
The two-qubit Trotterization gates in the considered case are given by
\[
G^\text{Heis} = e^{ -i\Delta t \left( \frac{h_1}{2}XI + \frac{h_2}{2}IX + \sum_{\alpha=1}^3 J^\alpha \sigma_i^\alpha \sigma_{i+1}^\alpha \right)},
\]
where 
\begin{itemize}[label=$\circ$]
    \item $h_1 = \frac{3}{4} = h_2$ for quantum gates that do not act on any edge qubits,
    \item $h_2 = \frac{3}{4} = \frac{h_1}{2}$ for quantum gates that act on the first qubit, and
    \item $h_1 = h = \frac{h_2}{2}$ for quantum gates that act on the last qubit
\end{itemize}
due to the open boundary conditions. 

\begin{figure}
    \centering
    \includegraphics[width=0.49\textwidth]{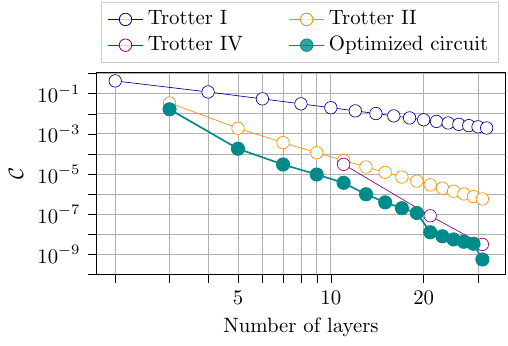}
    \caption{
    Results for the Heisenberg model on a chain with $N=50$ sites, $\vec{J}=(1,1,-\frac{1}{2})$, $\vec{h}=(\frac{3}{4},0,0)$, and $t=0.25$.
    }
    \label{fig:50_heisenberg}
\end{figure}

We optimize various quantum circuits that are initialized as the most accurate Trotterization for the considered number of layers. 
In our case, these are Trotter circuits of orders two and four as well as concatenations of both. 
The numerical results are presented in \cref{fig:50_heisenberg}. 
While the error improvement is not as significant as for the previous transverse-field Ising model, the Riemannian optimization increases the accuracy of all optimized quantum circuits to up to one order of magnitude. 

\subsection{Fermionic systems \label{subsec:fermionic-systems}}
When considering fermionic systems, their anti-symmetrical nature needs to be taken into account, e.g., utilizing second quantization. 
This study examines two types of fermionic Hamiltonians formulated in second quantization. 
A suitable fermion-to-qubit mapping such as the Jordan-Wigner transformation~\cite{duck1997pauli,Jordan:1928wi} must be applied to map the corresponding operators. 
Furthermore, fermionic swaps must be considered when simulating longer-range interactions using only nearest-neighbor two-qubit gates. 
Ref.~\cite{kivlichan2018quantum} has introduced the fermionic swap network to simulate a Trotter step of a fermionic system using a brickwall circuit layout. 
In the following, we introduce the fermionic systems we considered for our numerical simulations and provide further details on the Trotter implementation in \cref{ap:fermionic-swap-network}. 

\subsubsection{Spinful Fermi-Hubbard model.}
\begin{figure}
    \centering
    \includegraphics[width=0.49\textwidth]{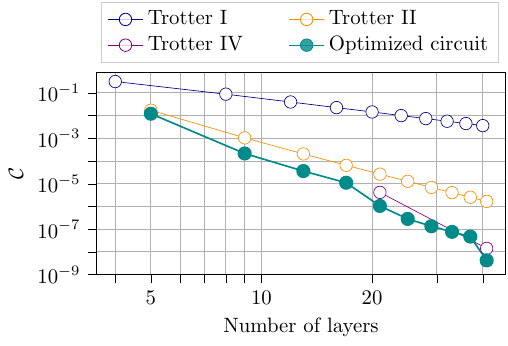}
    \caption{
    Results for the one-dimensional spinful Fermi-Hubbard model with $N=50$ spin orbitals, $T=1$, $V=4$, and $t=0.3$.
    }
    \label{fig:fermi-hubbard-results}
\end{figure}

As a first fermionic system, we consider the one-dimensional spinful Fermi-Hubbard model given by
\[
H^\text{FH} = -\sum_{\langle pq\rangle,s} T \left( a_{ps}^\dagger a_{qs} + a_{qs}^\dagger a_{ps} \right) + \frac{1}{2} \sum_p V n_{p\uparrow} n_{p\downarrow},
\]
where $\langle pq\rangle$ denotes the pairs of adjacent spatial orbitals on a chain, and $s\in\{\uparrow,\downarrow\}$ denotes the spin. 
Within the numerical simulations, we consider $T=1$ and $V=4$. 
We compute $\Uref$ as 10 repetitions of a fourth-order Trotterization, $\Trotter{IV}{,10}$, for a simulation time of $t=0.3$, which is further compressed to a threshold of $\errorThres$ as listed in \cref{ap:reference-error} \cref{tab:reference-overview}. 
To implement a Trotter step, a fermionic swap network is required, for which we provide more details in \cref{ap:fermionic-swap-network}. 
We start from quantum circuits initialized in Trotter steps of the second or fourth order as well as their concatenations, where the best Trotterization is chosen for a given number of layers. 
The optimization results are visualized in \cref{fig:fermi-hubbard-results}. 

Again, the Riemannian optimization method can improve the simulation accuracy for each optimized quantum circuit. 
The presented results show a relative error improvement of up to a factor of 6. 

\subsubsection{Molecular Hamiltonian}
\begin{figure}
    \centering
    \includegraphics[width=0.45\textwidth]{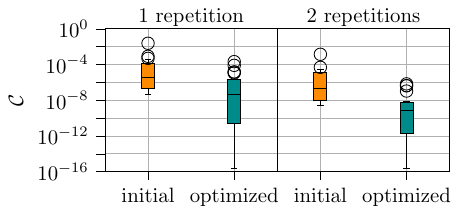}
    \caption{Riemannian optimization results for \ac{LiH}. The optimization was performed on 21 diagonal molecular Hamiltonian terms. For each term, a brickwall circuit initialized as one repetition (left panel) or two repetitions (right panel) of a second-order Trotter step is optimized.}
    \label{fig:LiH_results}
\end{figure}

Finally, we showcase a proof of concept on how the presented method could be applied to a chemical system. 
The general molecular Hamiltonian in second quantization is given by
\[
H^\text{mol} = \underbrace{\sum_{p,q = 1}^N t_{pq} \, a_p^\dagger a_q}_T + \underbrace{\sum_{p,q,r,s = 1}^N v_{pqrs} \, a_p^\dagger a_q^\dagger a_{r} a_s}_V,
\]
where the scalar coefficients $h_{pq}$ and $h_{pqrs}$ are the one- and two-electron integrals. 
Various techniques can be employed to ``diagonalize'' the Coulomb interaction term $V$~\cite{babbush2018low,white2017hybrid,luo2024efficient,motta2021low}. 
For the presented numerical experiment, we utilize the so-called \textit{double factorization}~\cite{motta2021low}, within which the two-body interaction term can be decomposed into diagonal terms via
\begin{align*}
    V &= \sum_{p,q} S_{pq} a_p^\dagger a_q + \sum_{\ell=1}^{N_\text{rot}} \sum_{i,j}^{\rho_\ell} \frac{\lambda_i^{(\ell)} \lambda_j^{(\ell)}}{2} n_i^{(\ell)} n_j^{(\ell)},
\end{align*}
where 
\[
n_i^{(\ell)} = \sum_{p,s = 1}^N U_{si}^{(\ell)} a_p^\dagger a_s U_{si}^{(\ell)} = a_{\psi_i^{(\ell)}}^\dagger a_{\psi_i^{(\ell)}}
\]
corresponds to the number operator in a rotated basis, $\psi_i^{(\ell)} = \sum_{p=1}^N U_{pi}^{(\ell)} \phi_p$. 
By doing so, $H^\text{mol}$ can be decomposed into Givens rotations and $N_\text{rot}$ Hamiltonian terms
\[
H_\text{diag}^{\text{mol}, (\ell)} = \sum_{i,j} T_{ij}^{(\ell)} a_{\psi_i^{(\ell)}}^\dagger a_{\psi_j^{(\ell)}} + V_{ij}^{(\ell)}n_i^{(\ell)} n_j^{(\ell)},
\]
where the two-body interaction $V^{(\ell)}$ is diagonal. 
A Trotter step for each $H_\text{diag}^{\text{mol},(\ell)}$ can be implemented following the fermionic swap network introduced in Ref.~\cite{kivlichan2018quantum}. 
Again, we leave further details on the fermionic swap network in \cref{ap:fermionic-swap-network}.
As a proof of concept, we consider the molecule \ac{LiH} with six orbitals and a simulation time $t=1$. 
We decompose its molecular Hamiltonian into $N_\text{rot}=21$ sets of rotations and diagonal Hamiltonians $\bigl\{H_{\text{diag}}^{\text{mol},(\ell)}\bigr\}_{\ell=1}^{N_\text{rot}}$, of which we optimize each. 
We start from an initial quantum circuit corresponding to one and two Trotter steps of order 2. 
The optimization results are visualized in \cref{fig:LiH_results}. 

We observe that the simulation accuracy could be increased for each considered quantum circuit. 
In the best cases, the optimized errors become negligible on the order of numerical precision ($\approx 10^{-16}$). 
However, we do not claim any generalization of these results to larger systems, as the considered molecule is a simple example intended to serve as a proof of concept. 
In each optimization case, a relative error improvement of at least one order of magnitude could be achieved, with the best relative error improvement being eight orders of magnitude. 
The median relative error improvement is about two orders of magnitude.

\section{Conclusion}
In this work, we present a novel approach that combines Riemannian optimization and tensor network methods to significantly enhance the simulation accuracy of initial Trotter circuits for the time evolution of various quantum systems. 
Our technique is purely classical, eliminating the need for costly quantum resources. 
To address the challenge of dimensionality in larger systems, we employ an \ac{MPO} approximation for the reference time evolution operator. 
This requires careful selection of the maximum bond dimension and simulation time to ensure computational feasibility while maintaining accurate \ac{MPO} representation of the actual time evolution. 
Our method is versatile and does not impose any symmetry constraints on the quantum systems. 
We demonstrate its effectiveness by optimizing ten disordered transverse-field Ising model instances. Additionally, we achieve error improvements of up to four orders of magnitude for one-dimensional systems such as the non-disordered transverse-field Ising model, the Heisenberg model, and the spinful Fermi-Hubbard model. 
We further showcase the applicability of our method to molecular Hamiltonians, using \ac{LiH} as a proof of concept, where we achieve error improvements of up to eight orders of magnitude. 

To tackle larger molecular systems, efficient methods to approximate the time evolution operator as an \ac{MPO} are crucial~\cite{paeckel2019time,van2024efficient}. 
We leave a more detailed analysis of this aspect for future work. 
Our current method employs first-order Riemannian optimization. 
However, recent results~\cite{Putterer2025riemannian} have shown a tremendous advantage of using second-order Riemannian optimization over first-order methods in the context of quantum circuit compression. 
Therefore, in future research, we plan to explore second-order optimization methods for brickwall circuits, such as the trust-region algorithm~\cite{kotil2024riemannian,Putterer2025riemannian}. 

\begin{acknowledgments}
We thank Joe Gibbs for valuable discussions about his recent work in Ref.~\cite{gibbs2024deep}.
We further thank Yu Wang for his input on tensor network methods. 
The Riemannian ADAM optimizer was implemented based on the ADAM implementation of \texttt{qiskit v0.18}~\cite{qiskit2024}. 
All tensor network operations were implemented using \texttt{jax.numpy}~\cite{jax2018github}. 
S.S. acknowledges support from the BMW group. 
The authors gratefully acknowledge the computational and data resources provided by the Leibniz Supercomputing Centre (\url{www.lrz.de}). 

\subsection*{Data availability}
The implementation used in this project, as well as the numerical results and a script to generate the figures in this publication, are available under \url{https://github.com/INMLe/rqcopt-mpo}. 

\subsection*{Authors contribution}
I.L. implemented the tensor network and optimization framework and performed the numerical simulations. 
S.S. implemented the double factorization of the molecular Hamiltonian. 
I.L. wrote the manuscript with contributions from all authors. 
C.M. supervised the project.

\end{acknowledgments}

\begin{acronym}
    \acro{SVD}{singular value decomposition}
    \acro{MPO}{matrix product operator}
    \acro{TEBD}{time-evolving block-decimation}
    \acro{LiH}{lithium hydride}
\end{acronym}
\bibliographystyle{quantum}
\bibliography{bibliography.bib}

\clearpage
\appendix

\setcounter{page}{1}
\renewcommand\thefigure{\thesection\arabic{figure}}
\setcounter{figure}{0} 

\onecolumngrid

\begin{center}
\large{ Supplementary Material for \\ ``Riemannian quantum circuit optimization based on matrix product operators''
}
\end{center}

\section{Gradient computation using the Wirtinger formalism\label{ap:gradient-computation}}
Within this work, we compute the gradient using the Wirtinger formalism~\cite{koor2023short} that can also be applied to non-holomorphic functions. 
In the following, we give a short summary of the latter and provide further details on the gradient calculation. 

Consider complex-valued smooth functions $f,g: \C \to \C$ and a complex number $z = x + i y$ with $x, y \in \R$.
The Wirtinger derivatives are defined as follows: 
\begin{subequations}
\begin{align}
\partial_z f(z)     &= \frac{1}{2}\left(\partial_x f - i \partial_y f \right), \\
\partial_{z^*} f(z) &= \frac{1}{2}\left(\partial_x f + i \partial_y f \right),
\end{align}
\end{subequations}
where $\partial_x f$ and $\partial_y f$ are the conventional partial derivatives when interpreting $f$ as function $f: \R^2 \to \C$, $f(x, y) = f(x + i y)$. 
Furthermore, the following \emph{chain rule} holds:
\begin{equation}
\label{eq:chain_rule}
\partial_z (f \circ g) = \big( \partial_w f \circ g \big) \partial_z g + \big( \partial_{w^*} f \circ g \big) \partial_z g^*,
\end{equation}
where ``$\circ$'' denotes function composition and $w = g(z)$. 
By definition, we can express the gradient $\grad f$ as 
\begin{equation}
\label{eq:gradient_Wirtinger_relation}
\grad f(z) = 2 \big(\partial_z f(z) \big)^\ast,
\end{equation}
where $(^\ast)$ denotes the complex conjugate.

In the following, we will give a more detailed computation of the Euclidean gradient $\grad \loss$. 
By using the presented Wirtinger formalism, the partial derivatives with respect to an entry of $W$ are given by
\[
    \partial _{W_{jk}} \Re \overlap = \frac{1}{2} \partial_{W_{jk}} \Tr(U^\dagger W + W^\dagger U) = \frac{U^\ast_{jk}}{2},
\]
and
\[
    \partial_{W_{jk}}\Im \overlap
    = \frac{1}{2i} \partial_{W_{jk}} \Tr(U^\dagger W - W^\dagger U) = \frac{U^\ast_{jk}}{2i},
\]
where we have used that 
\begin{align*}
\partial_{B_{\gamma\tau}}\Tr(A^\dagger B) = \sum_{\alpha,\beta} A_{\beta\alpha}^\ast B_{\beta\alpha} \delta_{\beta\gamma}\delta_{\alpha\tau} = A_{\gamma,\tau}^\ast,
\end{align*}
and 
\begin{align*}
\partial_{B_{\gamma\tau}}\Tr(B^\dagger A) = \partial_{B_{\gamma\tau}}\sum_{\alpha,\beta} B_{\beta\alpha}^\ast A_{\beta\alpha} = 0.
\end{align*}
For convenience, we repeat the loss function given in the main text in \cref{eq:lossHS}:
\[
\mc{L}=-\left[\Re\overlap\right]^2 - \left[ \Im\overlap \right]^2
\]
Using the chain rule given by \cref{eq:chain_rule} and the previous results, the partial derivative of $\mc{L}$ with respect to a matrix element $W_{jk}$ can be computed as
\begin{align*}
    \partial_{W_{jk}}\mc{L} &= -2\Re\overlap \cdot \partial_{W_{jk}} \Re\overlap \\
    &\;\;\;\;- 2\Im\overlap \cdot \partial_{W_{jk}} \Im\overlap\\
    &=  -\Re\overlap \cdot U_{jk}^\ast + \Im\overlap \cdot i U_{jk}^\ast \\
    &= -\left[\overlap U_{jk}\right]^\ast.
\end{align*}
By again applying the chain rule given in \cref{eq:chain_rule}, the partial derivatives of $\mc{L}$ with respect to a specific gate are given by
\begin{align*}
    \partial_{G_i^\ell} \mc{L} &= \partial_{W} \mc{L} \,\partial_{G_i^\ell} W = \sum_{jk} \partial_{W_{jk}} \mc{L} \cdot \partial_{G_i^\ell} W_{jk}\\
    &= - \sum_{jk} \overlap^\ast U_{kj}^\dagger\, \partial_{G_i^\ell} W_{jk}\\
    &= -\overlap^\ast\, \partial_{G_i^\ell} \sum_{jk}U_{kj}^\dagger W_{jk}\\
    &= -\overlap^\ast \cdot \partial_{G_i^\ell}\overlap,
\end{align*}
where we denote the number of layers in the brickwall circuit with $L$ and the number of gates in layer $\ell$ with $n_\ell$. 
Finally, the overall gradient can be obtained via \cref{eq:gradient_Wirtinger_relation} as
\begin{equation}
\grad\mc{L} = -2\,\overlap \cdot \partial_{G_i^\ell}\left[\overlap\right]_{\substack{\ell=1,\dots,L\\i=1,\dots,n_\ell}}^\ast\,.
\end{equation}

\section{Jacobians of the Weyl decomposition\label{ap:weyl-derivatives}}
For convenience, we state the Weyl decomposition~\cite{kraus2001optimal} again:
\[
G = (K_1 \otimes K_2) \cdot V(a,b,c) \cdot (K_3 \otimes K_4),
\]
with
\[
K_i = R_Z(\theta_i)\, R_Y(\psi_i)\, R_Z(\phi_i),
\]
and 
\[
V(a,b,c) = e^{i(aXX + bYY + cZZ)}
\]

The Jacobian with respect to $a,b,c\in\mathds{R}$ is given by:

\begin{align*}
    \frac{\partial G}{\partial a} &= iK_1\otimes K_2 \cdot XX \cdot V(a,b,c) \cdot K_3 \otimes K_3 \\
    \frac{\partial G}{\partial b} &= iK_1\otimes K_2 \cdot YY \cdot V(a,b,c) \cdot K_3 \otimes K_4 \\
    \frac{\partial G}{\partial c} &= iK_1\otimes K_2\cdot ZZ \cdot V(a,b,c) \cdot K_3 \otimes K_4.
\end{align*}

 The Jacobian of $K_j$ with respect to the corresponding angles $\theta_j, \psi_j, \phi_j \in\mathds{R}$ is given by:
 \begin{align*}
    \frac{\partial K_j}{\partial \theta_i} &= -\frac{i}{2} Z K_j \\
    \frac{\partial K_j}{\partial \psi_j} &= -\frac{i}{2} R_Z(\theta_j) Y  R_Y(\psi_j) R_Z(\psi_j) \\
    \frac{\partial K_j}{\partial \phi_j} &= -\frac{i}{2} R_Z(\theta_j) R_Y(\psi_j) Z R_Z(\phi_j).
\end{align*}
The Jacobian $\frac{\partial G}{\partial\alpha}$ for $\alpha=\theta_j,\psi_j,\phi_j$ can directly be obtained from the latter equations.

\section{Overview of reference approximation for the different Hamiltonians\label{ap:reference-error}}
Several error sources lead to a deviation of the reference \ac{MPO} and the exact time evolution operator as described in \cref{subsec:MPO-reference}. 
For the numerical simulations conducted in this work, we summarize these contributions in \cref{tab:reference-overview}. 
\begin{table*}
\caption{\label{tab:reference-overview}
Overview of reference approximation for the different Hamiltonians. 
}
\begin{tabular}{lcccc}
 & \textit{Reference approximation} & \textit{Trotter error} & \textit{Trunc. error} & \textit{Compr. error} \\
\hline
\multicolumn{1}{l}{\textit{Ising (20 qubits)}} & $\Trotter{IV}{,20}\;\;\;(t=2.0)$ & $1\cdot 10^{-11}$ & $5\cdot 10^{-15}$ & $9\cdot 10^{-8}$ \\
\multicolumn{1}{l}{\textit{Ising (50 qubits)}} & $\Trotter{IV}{,20}\;\;\;(t=2.0)$ & $4\cdot 10^{-11}$ & $2\cdot 10^{-14}$ & $2\cdot 10^{-9}$ \\
\multicolumn{1}{l}{\textit{Heisenberg (50 qubits)}}  & $\Trotter{IV}{,10}\;\;\;(t=0.25)$ & $4\cdot 10^{-11}$ & $3\cdot 10^{-14}$ & $3\cdot 10^{-11}$ \\
\multicolumn{1}{l}{\textit{Fermi-Hubbard (20 qubits)}}  & $\Trotter{IV}{,10}\;\;\;(t=0.3)$  & $5\cdot 10^{-11}$ & $2\cdot 10^{-11}$ & $5\cdot 10^{-11}$ \\
\multicolumn{1}{l}{\textit{Fermi-Hubbard (50 qubits)}}  & $\Trotter{IV}{,10}\;\;\;(t=0.3)$  & $2\cdot 10^{-10}$ & $5\cdot 10^{-11}$ & $5\cdot 10^{-11}$ \\
\end{tabular}
\end{table*}

\section{Fermionic swap network\label{ap:fermionic-swap-network}}
In this work, Trotter circuits of first-, second-, and fourth-order are considered.
Given a Hamiltonian of the form $H=\sum_{i=1}^k H_i$, these Trotterizations are given by~\cite{kluber2023trotterization}
\begin{align*}
    U_1(t) &= \prod_{i=1}^{k} e^{-i H_i t}, \\
    U_2(t) &= \prod_{i=1}^{k} e^{-i H_i t/2} \prod_{i=k}^{1} e^{-i H_i t/2}, \\
    U_4(t) &= \left[ U_2(s_2 t) \right]^2 \left[ U_2((1-4s_2)t) \right] \left[ U_2(s_2 t) \right]^2,
\end{align*}
where in the latter $s_2 = (4 -4^{1/3})^{-1}$.

To implement a Trotter circuit for fermionic Hamiltonians, each orbital $p$ is represented by a qubit $i_p$ and the fermionic operators $a^\dagger, a, n$ need to be mapped to corresponding qubit operators~\cite{kivlichan2018quantum}.
Using the Jordan-Wigner transformation~\cite{Jordan:1928wi,duck1997pauli} for this purpose results in possibly long-range Pauli-$Z$ strings.
However, if the qubits representing the involved orbitals $p,q$ are adjacent, i.e., $i_q=i_p+1$, the time-evolution of a single summand in $V$ can be implemented by two-qubit gates.
To this end, a fermionic swap network architecture can be employed~\cite{kivlichan2018quantum}, where fermionic swaps 
\[
f_\text{swap} = \begin{pmatrix}
    1 & 0 & 0 & 0 \\
    0 & 0 & 1 & 0 \\
    0 & 1 & 0 & 0 \\
    0 & 0 & 0 & -1\\
\end{pmatrix}
\]
are utilized suitably.

\subsection{Spinful Fermi-Hubbard model}
For simplicity, we state the considered one-dimensional spinful Fermi-Hubbard Hamiltonian again:
\[
H^\text{FH} = -\sum_{\langle pq\rangle,s} T_{pq} \left( a_{ps}^\dagger a_{qs} + a_{qs}^\dagger a_{ps} \right) + \frac{1}{2} \sum_p V_p n_{p\uparrow} n_{p\downarrow}.
\]
As can be seen, interactions of spin orbitals $(ps)$ and $(qs)$ for $p=q+1$ as well as interactions between spin orbitals $(p\uparrow)$ and $(p\downarrow)$ need to be considered.
To simulate a Trotter step, we order the spin orbitals as follows (example of 5 spatial orbitals):
\begin{figure}[H]
    \centering
    \includegraphics[width=0.48\textwidth]{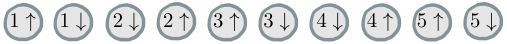}
\end{figure}

We then construct the swap network as follows:
The first and third layer simulate the kinetic hopping terms, for which the two-qubit gates $V_\text{kin}$ after Jordan-Wigner transformation are given by
\[
G_\text{kin}^\text{FH}(i_p,i_q) = \begin{pmatrix}
1 & 0 & 0 & 0 \\
0 & \cos(T_{pq} \Delta t) & i \sin(T_{pq} \Delta t) & 0 \\
0 & i \sin(T_{pq} \Delta t) & \cos(T_{pq} \Delta t) & 0 \\
0 & 0 & 0 & 1 \\
\end{pmatrix}.
\]
The second layer simulates the on-site interactions while simultaneously swapping the corresponding orbitals, for which the two-qubit gates $V_\text{int}$ after applying Jordan-Wigner transformation are given by
\[
G_\text{int}^\text{FH}(i_p,i_q) = \begin{pmatrix}
    1 & 0 & 0 &  0 \\
    0 & 0 & 1 &  0 \\
    0 & 1 & 0 &  0 \\
    0 & 0 & 0 & -e^{-iV_{pq}\Delta t}\\
\end{pmatrix}.
\]
Note that for first-order Trotterization, an additional layer that reverses the previous fermionic swaps given by needs to applied, which is not necessary for even-order Trotterizations.
The fermionic swap network for a first-order Trotter step of a system with 10 spin orbitals is then given by:
\begin{figure}[H]
    \centering
    \includegraphics[width=0.49\textwidth]{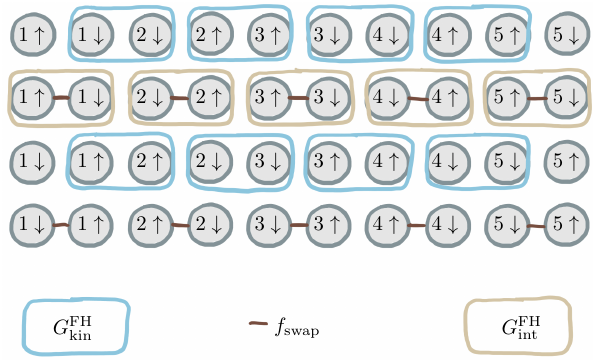}
\end{figure}

\subsection{Molecular Hamiltonian}
Ref.~\cite{kivlichan2018quantum} presented a fermionic swap network to implement a Trotter step of a molecular Hamiltonian with diagonal interaction term given by
\begin{equation*}
    H = \sum_{pq} T_{pq} a_p^\dagger a_q + \frac{1}{2}\sum_{p\neq q} V_{pq} n_p n_q,
\end{equation*}
where $p,q$ denote the orbitals.
In this case, the interaction and kinetics between two neighboring orbitals, as well as their fermionic swapping, can be jointly simulated by the following two-qubit gate:
\[
f_\text{sim}(i_p,i_q) = \begin{pmatrix}
1 & 0 & 0 & 0 \\
0 & -i\sin(T_{pq} \Delta t) & \cos(T_{pq} \Delta t) & 0 \\
0 & \cos(T_{pq} \Delta t) & -i\sin(T_{pq} \Delta t) & 0 \\
0 & 0 & 0 & -e^{-iV_{pq}\Delta t} \\
\end{pmatrix}
\]
The following fermionic swap network implements a Trotter step of first-order for 6 orbitals, where the layers for reversing the orbital ordering are neglected for simplicity:
\begin{figure}[H]
    \centering
    \includegraphics{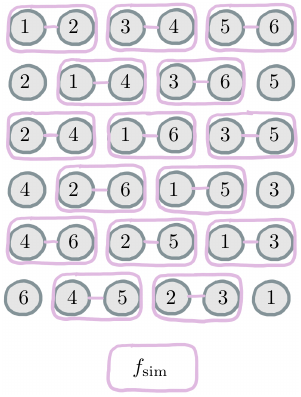}
\end{figure}

\section{Comparison of local and simultaneous gate updates\label{ap:comparison}}

Instead of updating all quantum gates simultaneously within a single optimization step, alternative ``sweeping'' approaches update one gate at a time while keeping the remaining quantum circuit fixed~\cite{causer2024scalable,gibbs2024deep}. 
In both methods, $\partial_{G_i^\ell} \overlap$ must be evaluated, as illustrated in \cref{fig:gradient-computation}. 
These partial derivatives contribute to the full gradient in Riemannian optimization. In the sweeping method, however, a specific gate $G_i^\ell$ is updated as 
\[
\Tilde{G}_i^\ell = q_\text{polar}\left[\partial_{G_i^\ell} \overlap^\dagger\right],
\] 
where $q_\text{polar}(A)$ denotes the unitary matrix $Q \in \mc{U}$ obtained from the polar decomposition $A = QP$, with $P \in \C^{4\times 4}$ being a Hermitian positive semi-definite matrix.
There are additional aspects in which the two optimization methods differ. 
In the local update method, there is flexibility in the order of sweeping through the quantum gates. 
It has been observed that different sweeping orders impact the optimization~\cite{causer2024scalable}; however, depending on the chosen local optimization method, not all sweeping orders can be implemented efficiently~\cite{gibbs2024deep}. 
Riemannian optimization updates all quantum gates simultaneously, thereby avoiding potential issues arising from a non-optimal sweeping order. 
However, Riemannian optimization requires an appropriate selection of the learning rate, which is not necessary for local updates. 
While the ADAM optimizer adjusts the learning rate during the optimization process, an appropriate initial choice can influence the quality of the optimization.

\begin{figure}
    \centering
    \includegraphics[width=0.35\linewidth]{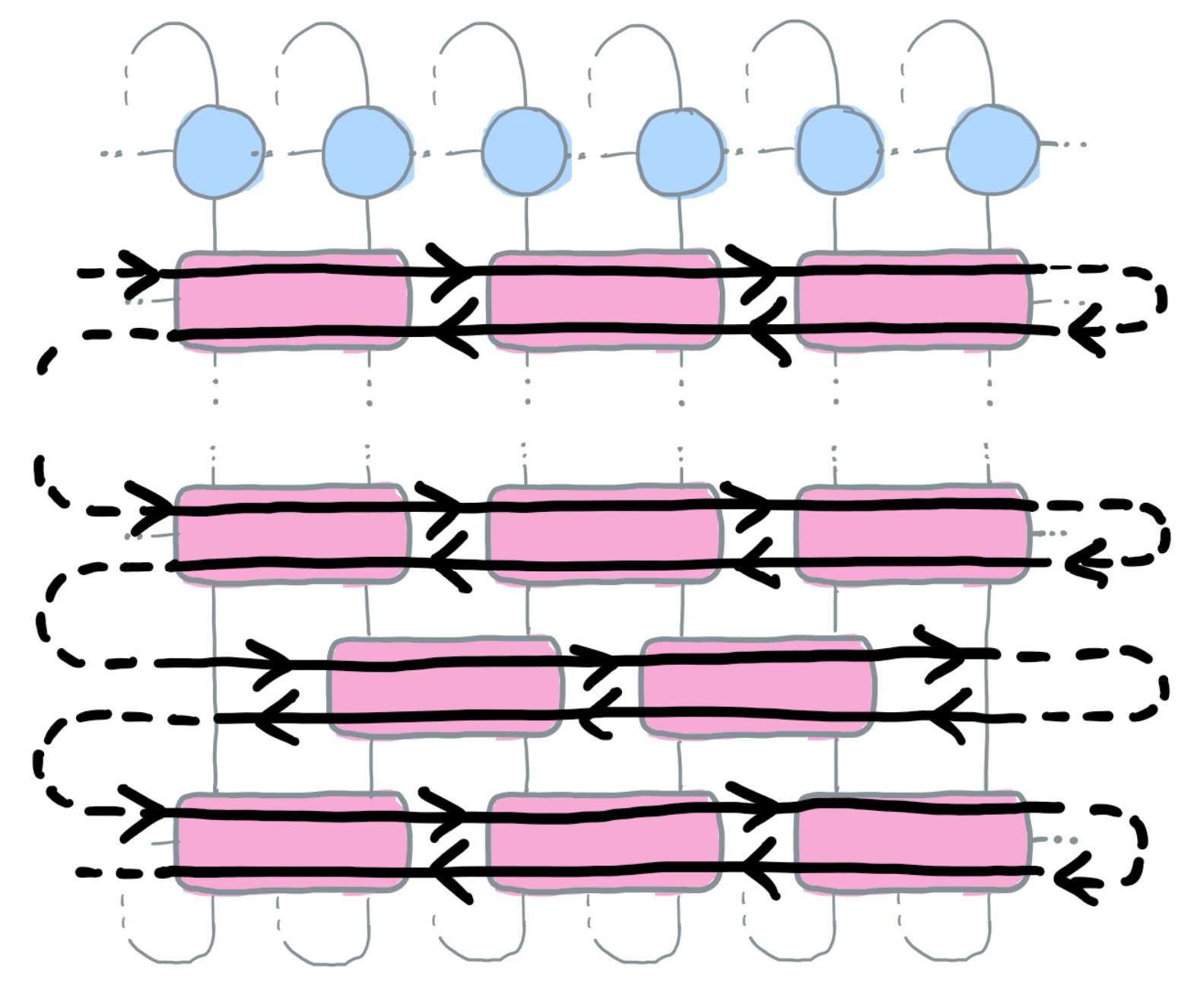}
    \caption{The chosen top-to-bottom sweeping order that defines one optimization step within the local update method. Similarly, a bottom-to-top sweep defines another optimization step. After one optimization step, each quantum gate was updated twice.}
    \label{fig:sweeping-order}
\end{figure}

To compare both optimization methods, we define one optimization step of the local sweep method as a top-to-bottom or bottom-to-top sweep, where each layer's quantum gates are updated in a back-and-forth manner, i.e., left-to-right-to-left. 
The corresponding sweeping order is illustrated in \cref{fig:sweeping-order}. 
This results in each quantum gate being updated twice per step, in contrast to the single update within the Riemannian approach. 
From a computational perspective, the bottleneck for Riemannian optimization is the computation of the gradient, whereas for the sweeping method, it is the local updates. 
Since both computations utilize a similar environment caching, the asymptotic computational scaling is the same.
\begin{figure}[b]
    \centering
    \includegraphics[width=0.49\textwidth]{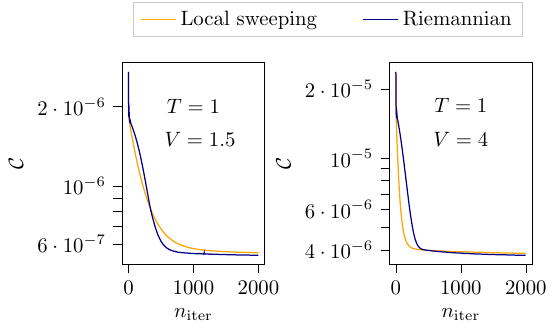}
    \caption{Loss curve comparison for a brickwall circuit with 17 layers. The considered systems are spinful Fermi-Hubbard models on 20 qubits with different interaction strengths.}
    \label{fig:comparison-loss}
\end{figure}
In the following, we consider two stopping criteria for the optimization: (i)~maximum number of iterations and (ii)~early stopping. 
In the first case, the optimization is terminated after a maximum number of iterations $n_\text{iter}$, and the optimized error for both methods is compared.
In the second case, the optimization is halted if the relative deviation of the optimization cost between two steps $i$ and $i-n$ is smaller than the tolerance given by
\[
2\cdot\left|\frac{\cost_{i-n} - \cost_{i}}{\cost_{i-n} + \cost_{i}}\right| \leq 10^{-5},
\]
where the distance between the two compared optimization steps is adjusted with the number of iterations, $n=\lceil 0.01i\rceil$.
For example, if the optimizer has proceeded for fewer than 100 iterations, the last two points are compared; if it has proceeded for 1000 steps, the last and the tenth last points are compared. 
If one method does not converge within $n_\text{iter}$ steps, we determine the number of iterations at which the convergence error of the other method is found. 
This is indicated with $(^\ast)$. 
We compare the sweeping method with the Riemannian method following the described procedure for two systems of a spinful Fermi-Hubbard chain on 20 qubits with interaction strengths $V=1.5$ and $V=4$ and $T=1$. 
Our findings are documented in \cref{tab:comparison-1}.
\begin{table*}
\caption{\label{tab:comparison-1} Convergence comparison between local sweeping method and Riemannian optimization for a spinful Fermi-Hubbard system on a chain of 20 qubits. $(^\ddag)$ indicates that the convergence criteria was not met within $n_\text{iter,max}$. In these cases, $n_\text{iter}$ was identified as the number of steps required to obtain the convergence error of the other method.}
\begin{tabular}{c|cc|cc||ccc}
\multicolumn{8}{c}{$T=1$ and $V=1.5$}\\\hline
$n_\text{layers}$ & \multicolumn{2}{c}{\makecell{\textit{Local} \\\textit{(early stopping)}}} & \multicolumn{2}{c||}{\makecell{\textit{Riemannian}\\\textit{(early stopping)}}} &  & \makecell{\textit{Local} \\\textit{(max. iteration)}} & \makecell{\textit{Riemannian}\\\textit{(max. iteration)}}\\\hline

 & \makecell{$n_\text{iter}$} & \textit{Error} & $n_\text{iter}$ & \textit{Error} & $n_\text{iter,max}$ & \textit{Error} & \textit{Error} \\\hline
    5     & $25$ &  $4.73\cdot 10^{-4}$ & $62$ & $4.73\cdot 10^{-4}$ & $100$ & $4.73\cdot 10^{-4}$ & $4.73\cdot 10{-4}$ \\
    9     & $1872^\ddag$ &  $1.09\cdot 10^{-5}$ & $1091$ & $1.09\cdot 10^{-5}$ & $2000$ & $1.09\cdot 10^{-5}$ & $1.81\cdot 10^{-5}$ \\
    14     & $1308^\ddag$ &  $1.87\cdot 10^{-6}$ & $799$ & $1.87\cdot 10^{-6}$ & $2000$ & $1.85\cdot 10^{-6}$ & $1.81\cdot 10^{-6}$ \\
    17     & -- &  -- & $914$ & $5.54\cdot 10^{-7}$ & $2000$ & $5.56\cdot 10^{-7}$ & $3.43\cdot 10^{-7}$ \\
    21     & $886$ &  $9.53\cdot 10^{-9}$ & $544$ & $9.53\cdot 10^{-9}$ & $2000$ & $9.13\cdot 10^{-9}$ & $8.85\cdot 10^{-9}$ \\
    41     & $310$ &  $7.04\cdot 10^{-11}$ & $817$ & $7.04\cdot 10^{-11}$ & $2000$ & $5.62\cdot 10^{-11}$ & $6.97\cdot 10^{-11}$ \\\hline\hline
\multicolumn{8}{c}{$T=1$ and $V=4$}\\\hline
$n_\text{layers}$ & \multicolumn{2}{c}{\makecell{\textit{Local} \\\textit{(early stopping)}}} & \multicolumn{2}{c||}{\makecell{\textit{Riemannian}\\\textit{(early stopping)}}} &  & \makecell{\textit{Local} \\\textit{(max. iteration)}} & \makecell{\textit{Riemannian}\\\textit{(max. iteration)}}\\\hline

 & \makecell{$n_\text{iter}$} & \textit{Error} & $n_\text{iter}$ & \textit{Error} & $n_\text{iter,max}$ & \textit{Error} & \textit{Error} \\\hline
    5     & $11$ &  $4.37\cdot 10^{-3}$ & $50$ & $4.37\cdot 10^{-3}$ & $100$ & $4.37\cdot 10^{-3}$ & $4.37\cdot 10{-3}$ \\
    9     & $560^\ddag$ &  $7.74\cdot 10^{-5}$ & $373$ & $7.74\cdot 10^{-5}$ & $2000$ & $7.73\cdot 10^{-5}$ & $7.73\cdot 10^{-5}$ \\
    14     & $1111^\ddag$ &  $1.28\cdot 10^{-5}$ & $577$ & $1.28\cdot 10^{-5}$ & $2000$ & $1.26\cdot 10^{-5}$ & $1.23\cdot 10^{-5}$ \\
    17     & $593^\ddag$ &  $3.96\cdot 10^{-6}$ & $541$ & $3.96\cdot 10^{-6}$ & $2000$ & $3.86\cdot 10^{-6}$ & $3.78\cdot 10^{-6}$ \\
    21     & $686^\ddag$ &  $3.13\cdot 10^{-7}$ & $518$ & $3.13\cdot 10^{-7}$ & $2000$ & $2.77\cdot 10^{-7}$ & $2.69\cdot 10^{-7}$ \\
    41     & $578$ &  $1.43\cdot 10^{-9}$ & $225$ & $1.43\cdot 10^{-9}$ & $2000$ & $1.29\cdot 10^{-9}$ & $1.33\cdot 10^{-9}$ \\
\end{tabular}
\end{table*}

We observe a generally similar convergence behavior with nearly identical final errors after $n_\text{iter,max}$ update steps and only a small variance in the required update steps when considering early stopping. 
An exemplary loss curve comparison for both Fermi-Hubbard models and a quantum circuit with 17 layers is depicted in \cref{fig:comparison-loss}.

Note that a one-to-one comparison of both methods is challenging since the local sweeping method generally does not have a natural global update step. 
Here, we chose a generous definition that allows the local sweeping method to update each gate twice per global update step, whereas one Riemannian optimization step updates each quantum gate only once. 
While both methods exhibit comparable optimization behavior, it is noteworthy that our method can, in principle, be extended to second-order Riemannian optimization~\cite{kotil2024riemannian,Putterer2025riemannian}, which requires a more computationally expensive evaluation of the Hessian but likely necessitates fewer optimization steps and is more robust against local minima.

\end{document}